\documentclass[preprint,prd,tightenlines,showpacs,nofootinbib,eqsecnum,amsfonts,amsmath,amssymb]{revtex4}

\usepackage{bm}
\usepackage{graphicx} 
\usepackage{hyperref}
\usepackage{mathrsfs}
\usepackage{multirow}    
\usepackage{dsfont}    
\usepackage{bbm}
\newcommand\id{\ensuremath{\mathbbm{1}}} 

\newcommand{\ud}{\mathrm{d}} 
\newcommand{\uD}{\mathrm{D}}
\newcommand{\calD}{\mathcal{D}}
\newcommand{\calO}{\mathcal{O}}

\newcommand{\ul}[1]{\underline{#1}{}}

\newcommand{\fond}[1]{\stackrel{\circ}{#1}\!{}}

\newcommand{\be}[1]{\begin{equation}\label{#1}}
\newcommand{\ee}{\end{equation}}
\newcommand{\bes}[1]{\begin{subequations}\label{#1}\begin{eqnarray}}
\newcommand{\ees}{\end{eqnarray}\end{subequations}}

\usepackage{ulem}
\usepackage[usenames]{color}

\allowdisplaybreaks

\begin{document}

\title{Phenomenology of Dark Matter via a \\ Bimetric Extension of
  General Relativity}

\author{Laura \textsc{Bernard}}\email{bernard@iap.fr}
\affiliation{$\mathcal{G}\mathbb{R}\varepsilon{\mathbb{C}}\mathcal{O}$
  Institut d'Astrophysique de Paris --- UMR 7095 du CNRS,
  \ Universit\'e Pierre \& Marie Curie, 98\textsuperscript{bis}
  boulevard Arago, 75014 Paris, France}

\author{Luc \textsc{Blanchet}}\email{blanchet@iap.fr}
\affiliation{$\mathcal{G}\mathbb{R}\varepsilon{\mathbb{C}}\mathcal{O}$
  Institut d'Astrophysique de Paris --- UMR 7095 du CNRS,
  \ Universit\'e Pierre \& Marie Curie, 98\textsuperscript{bis}
  boulevard Arago, 75014 Paris, France}

\date{\today}

\begin{abstract}
We propose a relativistic model of dark matter reproducing at once the
concordance cosmological model $\Lambda$-Cold-Dark-Matter
($\Lambda$-CDM) at cosmological scales, and the phenomenology of the
modified Newtonian dynamics (MOND) at galactic scales. To achieve this
we postulate a non-standard form of dark matter, consisting of two
different species of particles coupled to gravity \textit{via} a
bimetric extension of general relativity, and linked together through
an internal vector field (a ``graviphoton'') generated by the mass of
these particles. We prove that this dark matter behaves like ordinary
cold dark matter at the level of first order cosmological
perturbation, while a pure cosmological constant plays the role of
dark energy. The MOND equation emerges in the non-relativistic limit
through a mechanism of gravitational polarization of the dark matter
medium in the gravitational field of ordinary matter. Finally we show
that the model is viable in the solar system as it predicts the same
parametrized post-Newtonian parameters as general relativity.
\end{abstract}


\maketitle


\section{Introduction}
\label{Introduction}

The goal of the present article is to reproduce within a single
relativistic framework, consisting of a non-standard form of dark
matter particles coupled to a bimetric extension of general relativity
(GR), both:
\begin{enumerate}
\item the concordance cosmological model $\Lambda$-CDM and its
  tremendous successes at cosmological scales and notably the cosmic
  microwave background (CMB) at first order cosmological perturbations
  (see~\cite{OS95, HuD02, BHS05, Martin12} for reviews), in which cold
  dark matter (CDM) is a fluid of collisionless particles without
  interactions, and $\Lambda$ is a pure cosmological constant added to
  the Einstein field equations; and
\item the phenomenology of MOND (\textit{i.e.} MOdified Newtonian
  Dynamics or MilgrOmiaN Dynamics~\cite{Milg1, Milg2, Milg3}), which
  is a basic set of observational phenomena relevant to galaxy
  dynamics and dark matter distribution at galactic scales
  (see~\cite{SandMcG02, FamMcG12, McG14} for reviews), including most
  importantly the almost flat rotation curves of galaxies, the famous
  baryonic Tully-Fisher (BTF) relation for spiral galaxies~\cite{TF77,
    McG00, McG11}, and its equivalent for elliptical galaxies, the
  Faber-Jackson relation~\cite{Sand10}.
\end{enumerate}
It has long been known, and so far disappointing, that the
cosmological model $\Lambda$-CDM, when extrapolated down to galactic
scales, seems to be fundamentally incompatible with the
phenomenology of MOND. Within the $\Lambda$-CDM picture one can
only take notice of that phenomenology, and suppose that it emerges
from some (physical or astrophysical) mechanism taking place in the
interaction between dark matter and baryons. A lot of work on
astrophysical feedbacks (\textit{e.g.} supernova winds) has been done
to reconcile $\Lambda$-CDM with observations, see
\textit{e.g.}~\cite{SilkM12, SBCFT12}. However, because of the problem
of fine tuning of complicated phenomena to simple empirical laws like
the BTF relation, and because of the baffling presence of the MOND
acceleration scale $a_0$ in the data, it appears to be practically
impossible that $\Lambda$-CDM could provide a satisfactory explanation
of the MOND phenomenology~\cite{McG14}. By contrast, the MOND
empirical formula is extremely predictive and successful for galaxy
dynamics~\cite{SandMcG02, FamMcG12}, but is antagonistic to anything
we would like to call a fundamental theory. Furthermore, it has
problems at larger scales where it fails to reproduce about one half
of the dark matter we see in galaxy clusters~\cite{GD92, Sand99,
  PSilk05, Clowe06, Ang08, Ang09}, and unfortunately has \textit{a
  priori} little to say about cosmology at still larger scales.

Most relativistic MOND theories extend GR with appropriate extra
fundamental fields, so as to recover MOND in the non-relativistic
limit, see Refs.~\cite{Sand97, Bek04, Sand05, ZFS07, Halle08, bimond1,
  bimond2, BDgef11, BM11, Sand11}. None of these theories assume
dark matter, so they can be called pure modified gravity
theories. They have been extensively studied in cosmology, notably the
Tensor-Vector-Scalar (TeVeS) theory~\cite{Sand97, Bek04, Sand05} and
non-canonical Einstein-{\ae}ther theories~\cite{ZFS07, Halle08}, at
first order perturbation around a cosmological background (see
\textit{e.g.} Refs.~\cite{SMFB06, LMB08, LBMZ08, Sk06, Sk08,
  Zu10}). However, because they do not assume dark matter, the pure
modified gravity theories have difficulties at reproducing the
cosmological observations, notably the full spectrum of anisotropies
of the CMB.

A different approach, called dipolar dark matter (DDM), is more
promising in order to fit cosmological observations. This approach is
motivated by the \textit{dielectric analogy} of MOND~\cite{B07mond,
  BBwag}, a remarkable property of the MOND formula which may have
deep physical implications (but of course, which could also be merely
coincidental). The idea is that the phenomenology of MOND could arise
from some property of dark matter itself, namely a space-like vector
field called the gravitational dipole moment and able to polarize the
DDM medium in the gravitational field of ordinary matter. A
relativistic version of this idea has been proposed in
Refs.~\cite{BL08, BL09}, and correctly reproduces the cosmological
$\Lambda$-CDM model at the level of first order cosmological
perturbations. The deviations from $\Lambda$-CDM at second order
cosmological perturbations in that model have also been
investigated~\cite{BLLM13}.

In the model~\cite{BL08, BL09} the phenomenology of MOND is recovered
when the DDM medium is polarized, \textit{i.e.} when the polarization
field is aligned with the local gravitational field. This is obtained
at the price of a hypothesis of ``weak clustering'' of DDM, namely
the fact that the DDM medium stays essentially at rest and does not
cluster much in galaxies compared to ordinary matter. This hypothesis
is made plausible by the fact that the internal force due to the
presence of the dipole moment will balance the gravitational
force. Furthermore the hypothesis has been explicitly verified in the
case of the static gravitational field of a spherical mass
distribution (see the Appendix of Ref.~\cite{BL08}). However, in more
general situations, either highly dynamical or involving non-spherical
gravitational fields, it is likely that the polarization will not be
exactly aligned with the gravitational field, and in that case the
model~\cite{BL08, BL09} would deviate from MOND \textit{stricto
  sensu}. We have in mind situations like the dynamical evolution of
galaxies including the formation of bars~\cite{TC07}, the collision of
spiral galaxies yielding famous antenna structures~\cite{TC08} and the
formation of tidal dwarf galaxies~\cite{GFC07}, and the problem of
non-spherical polar ring galaxies~\cite{LFK13}.

In the present paper we propose a new relativistic model for DDM,
which is free of the weak clustering hypothesis of the DDM, and thus
permits one to recover the phenomenology of MOND in all situations, either
spherical or non-spherical, and/or highly dynamical. Furthermore we
shall show that this model also recovers the essential features of the
standard cosmological model $\Lambda$-CDM and in particular is
indistinguishable from it at first order perturbation around a
cosmological background.

The present model is actually closer to the original concept of
gravitational polarization and dipolar dark matter~\cite{B07mond,
  BBwag}. Indeed it involves \textit{two} species of dark matter
particles, interacting together \textit{via} some internal force
field. The DDM medium appears to be the gravitational analogue of a
plasma in electrodynamics, oscillating at the natural ``plasma''
frequency, and which can be polarized by the gravitational field of
ordinary matter, mimicking the presence of dark matter. The
non-relativistic approximation of our model has already been reviewed
in Sec.~III of Ref.~\cite{BBwag} and exhibits all the desirable
features we would expect for gravitational polarization and
MOND.\,\footnote{Interpreting this polarizable dark matter medium
    as a sea of virtual pairs of particles and antiparticles, we gave
    in Ref.~\cite{BBwag} a few numerical estimates that such an
    hypothetical medium would have~\cite{Chardin09,Hajdu11}.}

To achieve these results we assume that the two species of dark matter
particles are coupled to two different metrics, reducing to two
different Newtonian potentials in the non-relativistic
limit~\cite{BBwag}. Therefore, in this new model and contrary to the
previous one~\cite{BL08, BL09}, we do consider a modification of
gravity, in the form of a bimetric extension of GR. Furthermore, the
internal field is chosen to be a vector field whose associated charge
is the mass of particles --- \textit{i.e.} a ``graviphoton''. Thus our
model is a compromise between particle dark matter and modified
gravity, which can be seen as the result of the antinomic
phenomenologies of dark matter when it is seen either in cosmology or
in galaxies. Note that the model is very different from
BIMOND~\cite{bimond1, bimond2}, a bimetric theory that has been
proposed for MOND and which is a pure modified gravity theory without
dark matter.

As our model uses a bimetric extension of general relativity, it is
necessary to check the consistency of its gravitational sector. In
particular the number of propagating gravitational degrees of freedom
should be investigated, together with their possible ghost-like
behaviour. This work would be along the lines of studies of ghost-free
bimetric theory motivated by a non-trivial generalisation of de Rham, Gabadadze and Tolley (dRGT)
massive gravity~\cite{deRham10, dRGT10, Hassan11, HassanR11,
  BDvS14}. In the spirit of the search for relativistic MOND
theories~\cite{Sand97, Bek04, Sand05, ZFS07, Halle08, bimond1,
  bimond2, BDgef11, BM11, Sand11,BL08, BL09}, we focus here on the
physical consequences of the model. The counting of gravitational
  propagating modes is treated in Ref.~\cite{BH15}. However we shall
indicate in Appendix~\ref{app:lineargravact} that some aspects of the
gravitational sector of the model are safe at linear order around a
Minkowski background.

Finally since the present theory involves a modification of gravity it
is very important to check its viability in the Solar System (SS). We
compute the first post-Newtonian (1PN) limit of the model in the
regime of the SS, \textit{i.e.} when typical accelerations are much
above the MOND scale $a_0$, and find the same parametrized
post-Newtonian (PPN) parameters as in GR~\cite{Will}, which allows us
to conclude that the theory is viable in this regime.

The plan of this paper is as follows. In Sec.~\ref{sec:Model} we
describe the model using a relativistic action for the ordinary
matter, the two types of dark matter coupled to two different metrics,
and an internal vector field. We also look at the perturbative
solution of the field and matter equations. In Sec.~\ref{sec:cosmo} we
investigate the cosmology of the model up to first order in
perturbations.  In Sec.~\ref{sec:NR1PN} we investigate the
non-relativistic limit of the model, describe the mechanism of
polarization that yields the MOND phenomenology at galactic scales
(see~\cite{BBwag} for a review), and check the 1PN limit of the
model. The paper ends with a short conclusion in Sec.~\ref{sec:concl},
and with appendices presenting technical details, notably
Appendix~\ref{app:lineargravact} which investigates the
gravitational sector of the model at linear order.


\section{Dipolar dark matter and modified gravity}
\label{sec:Model}

\subsection{Dynamical action and field equations}
\label{sec:action}

Let us consider a model involving, in addition to the ordinary matter
simply described by baryons, two species of dark matter particles. The
gravitational sector is composed of two Lorentzian metrics
$g_{\mu\nu}$ and $\ul{g}_{\mu\nu}$ and one vector field $K_\mu$
sourced by the dark matter masses and which will be called a
\textit{graviphoton}. The baryons are coupled in the usual way to the
metric $g_{\mu\nu}$. Though we model the ordinary matter only by
baryons, we have in mind that all ordinary matter fields (fermions,
neutrinos, electromagnetic radiation, \textit{etc.}) are
coupled in the standard way to the ordinary metric $g_{\mu\nu}$. As a
way to recover the dipolar behaviour of dark matter we assume that one
species of dark matter particles is, like the baryons, minimally
coupled to the ordinary metric $g_{\mu\nu}$, while the other one is
minimally coupled to the second metric $\ul{g}_{\mu\nu}$. The vector
field $K_\mu$ links together the two species of dark matter particles
and is crucial in order to ensure the stability of the dipolar medium.

The gravitational-plus-matter action of our model
reads\,\footnote{Greek indices $\mu, \nu, \cdots$ take space-time
  values $0,1,2,3$ and Latin ones space values $1,2,3$. The signature
  of the three Lorentzian metrics $g_{\mu\nu}$, $\ul{g}_{\mu\nu}$ and
  $f_{\mu\nu}$ is $(-,+,+,+)$. In most of the paper we use geometrical
  units with $G=c=1$. Symmetrization of indices is defined by
  $T_{(\mu\nu)}=\frac{1}{2}(T_{\mu\nu}+T_{\nu\mu})$.}
\begin{align}\label{action} S &= \int\ud^{4}x\left\{
\sqrt{-g}\left(\frac{R-2\lambda}{32\pi}-\rho_\text{b}-\rho\right) +
\sqrt{-\ul{g}}\left(\frac{\ul{R}-2\ul{\lambda}}{32\pi} -
\ul{\rho}\right)\right. \nonumber\\ &
\left.\qquad\qquad+\sqrt{-f}\left[\frac{\mathcal{R}-2\lambda_f}{16\pi
    \varepsilon} +(j^\mu-\ul{j}^{\mu})K_\mu +
  \frac{a_{0}^{2}}{8\pi}\,W(X)\right] \right\} \,.\end{align}
We describe baryons and dark matter particles in their respective
sectors by their conserved scalar densities $\rho_\text{b}$, $\rho$
and $\ul{\rho}$, without pressure, and define their four velocities
$u_\text{b}^\mu$, $u^\mu$ and $\ul{u}^\mu$, normalized with their
respective metrics, \textit{i.e.} $g_{\mu\nu}u_\text{b}^\mu
u_\text{b}^\nu=g_{\mu\nu}u^\mu u^\nu=\ul{g}_{\mu\nu}\ul{u}^\mu
\ul{u}^\nu=-1$. We denote by $R\equiv R[g]$ and $\ul{R}\equiv
R[\ul{g}]$ the Ricci scalars associated with the two metrics
$g_{\mu\nu}$ and $\ul{g}_{\mu\nu}$. These metrics interact with each
other through an interaction term involving the Ricci scalar
$\mathcal{R}\equiv R[f]$ associated with an additional Lorentzian
metric $f_{\mu\nu}$, defined non-perturbatively from $g_{\mu\nu}$ and
$\ul{g}_{\mu\nu}$ by the implicit relations
\be{fdef} f_{\mu\nu} = f^{\rho\sigma} g_{\rho\mu}\,\ul{g}_{\nu\sigma}
= f^{\rho\sigma} g_{\rho\nu}\,\ul{g}_{\mu\sigma} \,,\ee
where $f^{\rho\sigma}$ is the inverse metric, \textit{i.e.}
$f^{\rho\sigma}f_{\sigma\tau}=\delta^{\rho}_\tau$. Note
that~\eqref{fdef} implies $f^2=g\,\ul{g}$ for the determinants
[\textit{e.g.} $f=\text{det}(f_{\mu\nu})$]. In applications the
relations~\eqref{fdef} will be solved perturbatively and the solution
in the form of a full perturbative series is analyzed in
Appendix~\ref{app:solpert}. The action~\eqref{action} is thus composed
of an ordinary sector coupled to $g_{\mu\nu}$, first term
in~\eqref{action}, a dark sector coupled to $\ul{g}_{\mu\nu}$ in the
second term, and an interacting sector with metric $f_{\mu\nu}$ in the
third term, which also entirely contains the contribution of the
internal field $K_\mu$. The ordinary and dark sectors are not
symmetrical due to the baryons in the ordinary sector, and we
may imagine that this is somewhat similar to the matter-antimatter
asymmetry.

The model is specified by several constants: the MOND acceleration
scale $a_0$~\cite{Milg1, Milg2, Milg3} which has been introduced only
in the interacting sector, some cosmological constants $\lambda$,
$\ul{\lambda}$ and $\lambda_f$ that have been inserted in the three
sectors and will be related to the true cosmological constant
$\Lambda$ of the model $\Lambda$-CDM, and a dimensionless coupling
constant $\varepsilon$ ruling the strength of the interaction between
the two metrics and which will be assumed to be very small,
$\varepsilon\ll 1$, in Sec.~\ref{sec:NR1PN}. We can write the latter
coupling constant as $\varepsilon=(m_\text{P}/M)^2$ where $m_\text{P}$
is the Planck mass and $M$ represents a new mass scale that we shall
not need to specify here except in Sec.~\ref{sec:NR1PN} where $M\gg
m_\text{P}$ will be assumed.

Like in the previous model~\cite{BL08, BL09} we have in mind that the
MOND scale $a_0$ is a fundamental constant, and that the observed
cosmological constant $\Lambda$ (to which we shall relate the
constants $\lambda$, $\ul{\lambda}$, $\lambda_f$ in the action) would
be derived from it in a more fundamental theory, and so naturally
satisfies the appropriate scaling relation $\Lambda\sim a_0^2$ which
is in very good agreement with observations~\cite{FamMcG12}. It would
be interesting to investigate whether the coupling constant
$\varepsilon$ (or mass $M$) could also be related to the
acceleration scale $a_0$.

The internal vector field $K_\mu$ obeys a non-canonical kinetic term
$W(X)$ where
\be{X} X\equiv -\frac{H^{\mu\nu} H_{\mu\nu}}{2a_{0}^{2}}\,. \ee
We pose $H_{\mu\nu}=\partial_\mu K_\nu-\partial_\nu K_\mu$ and
$H^{\mu\nu}=f^{\mu\rho}f^{\nu\sigma}H_{\rho\sigma}$ since the metric
in this sector is $f_{\mu\nu}$. Note that the vector field strength
in~\eqref{X} has been rescaled by the MOND acceleration $a_0$. We
refer to~\cite{EPU10} for discussions on the stability and Cauchy
problem for vector field theories involving non-canonical kinetic
terms.

The function $W$ is determined phenomenologically in order to recover
MOND from the non-relativistic limit of the model studied in
Sec.~\ref{sec:NRlimit}, and to be in agreement with the usual
solar-system tests as investigated in Sec.~\ref{sec:1PNlimit}. This
function, which is related to the MOND interpolating function, should
in principle be interpreted within some more fundamental
theory. However this task is not addressed in this work. In the limit
$X\ll 1$, which corresponds to the MOND weak-acceleration regime below
the scale $a_0$, we impose
\be{W0} W(X)= X-\frac{2}{3} X^{3/2} + \calO\left(X^2\right)\,.\ee
On the other hand we also impose the following behaviour of $W$ when
$X\gg 1$ so as to recover the usual 1PN limit of GR in an acceleration
regime much above $a_0$ [see Sec.~\ref{sec:1PNlimit}],
\be{Winf} W(X)= A + \frac{B}{X^{b}} +
o\!\left(\frac{1}{X^{b}}\right)\,,\ee
where $A$ and $B$ are constants and where the power $b$ can be any
strictly positive real number, $b>0$. The limit of the
action~\eqref{action} in the strong field regime $X\gg 1$ is
given in Eq.~\eqref{actionSS} of Sec.~\ref{sec:NR1PN}.

The graviphoton $K_\mu$ is sourced by the dark matter currents $j^\mu$
and $\ul{j}^\mu$ in the interacting sector of the action, defined as
follows. First we define the baryons and dark matter currents in their
respective sector by $J_\text{b}^\mu=\rho_\text{b} u_\text{b}^\mu$,
$J^\mu=\rho u^\mu$ and $\ul{J}^\mu=\ul{\rho} \ul{u}^\mu$. These
currents are conserved in the sense that $\nabla_\mu J_\text{b}^\mu=0$
and $\nabla_\mu J^\mu=0$, where $\nabla_\mu$ is the covariant
derivative associated with $g_{\mu\nu}$, and
$\ul{\nabla}_\mu\ul{J}^\mu=0$, where $\ul{\nabla}_\mu$ is the
covariant derivative of $\ul{g}_{\mu\nu}$. Then both dark matter
currents $j^\mu$ and $\ul{j}^\mu$ in the action~\eqref{action} are
defined with respect to the metric $f_{\mu\nu}$, solution of
Eq.~\eqref{fdef}. They are thus given by
\be{jmu} j^\mu = \beta\,J^\mu\,,\qquad \ul{j}^\mu =
\ul{\beta}\,\ul{J}^\mu\,,\ee
where we pose $\beta\equiv\sqrt{-g}/\sqrt{-f}$ and
$\ul{\beta}\equiv\sqrt{-\ul{g}}/\sqrt{-f}$,\,\footnote{Note that with
  our choice~\eqref{fdef} for the metric $f_{\mu\nu}$ we have
  $\beta\ul{\beta}=1$.} and obey the conservation laws $\calD_\mu
j^\mu=0$ and $\calD_\mu \ul{j}^\mu=0$, where $\calD_\mu$ is the
covariant derivative associated with $f_{\mu\nu}$.

As a preliminary check of the consistency of our model, we
investigate in Appendix~\ref{app:lineargravact} the gravitational
part of the action~\eqref{action} at quadratic order around a
Minkowski background, and show that it reduces to the sum of the
actions for two non-interacting massless spin-2 fields. We conclude
that the model is consistent (\textit{i.e.} ghost-free) at that
order.

First we vary the action with respect to the metrics $g_{\mu\nu}$ and
$\ul{g}_{\mu\nu}$. For the moment we write the linear variation of
$f_{\mu\nu}$ as
\be{deltaf} \delta f_{\mu\nu} =
\frac{1}{2}\,\mathcal{A}_{\,\mu\nu}^{\,\rho\sigma}\,\delta
g_{\rho\sigma} +
\frac{1}{2}\,\ul{\mathcal{A}}_{\,\mu\nu}^{\,\rho\sigma}\,\delta
\ul{g}_{\rho\sigma}\,,\ee
where $\mathcal{A}_{\,\mu\nu}^{\,\rho\sigma}$ and
$\ul{\mathcal{A}}_{\,\mu\nu}^{\,\rho\sigma}$ denote some tensorial
coefficients obeying two implicit equations given in
Appendix~\ref{app:solpert}, and which will be computed perturbatively
in applications, see for instance~\eqref{solAAbar2} in
Appendix~\ref{app:solpert}. We then obtain the two Einstein field
equations
\bes{EE12} \beta\Bigl(E^{\mu\nu} + \lambda \,g^{\mu\nu}\Bigr) +
\frac{1}{\varepsilon}\,\mathcal{A}^{\,\mu\nu}_{\,\rho\sigma}
\Bigl(\mathcal{E}^{\rho\sigma}+\lambda_f f^{\rho\sigma}\Bigr) &=&
16\pi\biggl[ \beta\bigl(T_\text{b}^{\mu\nu}+T^{\mu\nu}\bigr) +
  \mathcal{A}^{\,\mu\nu}_{\,\rho\sigma}\,\tau^{\rho\sigma}
  \biggr]\,,\\ \ul{\beta}\Bigl(\ul{E}^{\mu\nu} + \ul{\lambda}
\,\ul{g}^{\mu\nu}\Bigr) + \frac{1}{\varepsilon}
\,\ul{\mathcal{A}}^{\,\mu\nu}_{\,\rho\sigma}
\Bigl(\mathcal{E}^{\rho\sigma}+\lambda_f f^{\rho\sigma}\Bigr) &=&
16\pi\biggl[ \ul{\beta}\,\ul{T}^{\mu\nu} +
  \ul{\mathcal{A}}^{\,\mu\nu}_{\,\rho\sigma}\,\tau^{\rho\sigma}\biggr]\,,\ees
where the Einstein tensors associated with their respective metrics
are $E^{\mu\nu} = R^{\mu\nu} - \frac{1}{2}g^{\mu\nu}R$,
$\ul{E}^{\mu\nu} = \ul{R}^{\mu\nu} - \frac{1}{2}\ul{g}^{\mu\nu}\ul{R}$
and $\mathcal{E}^{\mu\nu} = \mathcal{R}^{\mu\nu} -
\frac{1}{2}f^{\mu\nu}\mathcal{R}$. The stress-energy tensors of the
matter particles are given by $T_\text{b}^{\mu\nu} =
\rho_\text{b}u_\text{b}^{\mu}u_\text{b}^{\nu}$, $T^{\mu\nu} = \rho
u^{\mu}u^{\nu}$ and $\ul{T}^{\mu\nu} =
\ul{\rho}\ul{u}^{\mu}\ul{u}^{\nu}$, each one being defined with its
respective metric. In addition, the stress-energy tensor of the
internal graviphoton field $K_\mu$, living in the sector $f_{\mu\nu}$,
reads
\be{taumunu} \tau^{\mu\nu} =
\frac{1}{8\pi}\biggl[W'\,H^{\mu\rho}H^{\nu}_{\phantom{\nu}\rho} +
  \frac{a_0^2}{2}\,W f^{\mu\nu}\biggr]\,, \ee
where $W'\equiv\ud W/\ud X$. Next, varying the action with respect to
$K_{\mu}$ we obtain
\be{graviphoton} \calD_{\nu}\Bigl[W'\,H^{\mu\nu}\Bigr] =
4\pi\left(j^{\mu}-\ul{j}^{\mu}\right)\,,\ee
which is obviously compatible with the conservation laws $\calD_\mu
j^\mu=\calD_\mu \ul{j}^\mu=0$.

Finally, we vary the action with respect to the particles. Since the
baryons are minimally coupled to the metric $g_{\mu\nu}$, their
equation of motion is simply the geodesic equation, $a_\text{b}^\mu=0$
where $a_\text{b}^\mu\equiv u_\text{b}^\nu\nabla_\nu
u_\text{b}^\mu$. On the contrary, because of the presence of the
internal field $K_\mu$, the motion of dark matter particles is
non-geodesic,
\bes{DMeom} a_{\mu} &=& u^{\nu}\,H_{\mu\nu} \,,\\ \ul{a}_{\mu} &=&
-\ul{u}^{\nu}\,H_{\mu\nu} \,,\ees
where $a^\mu\equiv u^\nu\nabla_\nu u^\mu$ and $a_\mu=g_{\mu\nu}a^\nu$,
and similarly $\ul{a}^\mu\equiv \ul{u}^\nu\ul{\nabla}_\nu \ul{u}^\mu$
and $\ul{a}_\mu=\ul{g}_{\mu\nu}\ul{a}^\nu$. Note that the forces
acting on the two species of dark matter particles are space-like, and
are completely analogous to the usual Lorentz force acting on charged
particles.

The stress-energy tensors of the dark matter particles and of the
internal field are not conserved separately, but we can derive a
``global'' conservation law. Indeed the equation~\eqref{graviphoton}
can be equivalently written by means of the stress-energy
tensor~\eqref{taumunu} as
\be{divtau} \calD_{\nu} \tau^\nu_\mu = - \frac{1}{2}\left(j^\nu -
\ul{j}^\nu \right)H_{\mu\nu}\,, \ee
where we pose $\tau^\nu_\mu=f_{\mu\rho}\tau^{\nu\rho}$. As a result of
Eq.~\eqref{divtau} the two dark matter equations of
motion~\eqref{DMeom} can be combined to give
\be{cons} \calD_\nu \tau^\nu_\mu +
\frac{1}{2}\Bigl(\beta\,\nabla_\nu
T^\nu_\mu+\ul{\beta}\,\ul{\nabla}_\nu
\ul{T}^\nu_\mu\Bigr) = 0\,, \ee
where $T^\nu_\mu=g_{\mu\rho}T^{\nu\rho}$ and
$\ul{T}^\nu_\mu=\ul{g}_{\mu\rho}\ul{T}^{\nu\rho}$. This conservation
law describes the exchanges of stress-energy between the dark matter
particles and the internal field.

\subsection{First order perturbation of the matter and gravitational fields}
\label{sec:displ}

We now make a crucial assumption regarding the two fluids of dark
matter particles, namely that they differ by some small displacement
vectors $y^\mu$ and $\ul{y}^\mu$ from a common equilibrium
configuration where they superpose on top of each other. This
assumption permits one to obtain a solution of the field equations, which
is at the basis of the cosmological, MOND and solar-system solutions,
respectively investigated in Secs.~\ref{sec:cosmo},~\ref{sec:NRlimit}
and~\ref{sec:1PNlimit}. Such a solution suggests a description of the
dark matter medium as the analogue of a relativistic plasma in
electromagnetism, polarizable in the gravitational field of ordinary
matter and oscillating at its natural plasma frequency~\cite{B07mond,
  BBwag}.

Looking for such a solution we make a perturbative assumption
regarding the two metrics $g_{\mu\nu}$ and $\ul{g}_{\mu\nu}$. We note
that if they are related by a conformal transformation,
$g_{\mu\nu}=\alpha^2\ul{g}_{\mu\nu}$, then there is a simple,
``conformal'' solution of~\eqref{fdef} given by
$f_{\mu\nu}=\alpha^{-1}g_{\mu\nu}=\alpha\ul{g}_{\mu\nu}$. Here we
assume that our solution differs from the latter conformal solution by
a small metric perturbation
$h_{\mu\nu}=\frac{1}{2}(\alpha^{-1}g_{\mu\nu}-\alpha\ul{g}_{\mu\nu})$.
Then we can solve Eq.~\eqref{fdef} at first order in $h_{\mu\nu}$ as
\bes{ggbar} g_{\mu\nu} &=& \alpha\bigl(f_{\mu\nu} + h_{\mu\nu}\bigr) +
\calO(2)\,,\\ \ul{g}_{\mu\nu} &=& \frac{1}{\alpha}\bigl(f_{\mu\nu} -
h_{\mu\nu}\bigr) + \calO(2)\,,\ees
where second-order terms in $h_{\mu\nu}$ are systematically neglected
in this section and we define $\calO(n)\equiv\calO(h^n)$. Our
introduction of the factor $\alpha$ is motivated by the application to
cosmology in Sec.~\ref{sec:cosmo} in order to allow for two different
cosmological backgrounds for the metrics $g_{\mu\nu}$ and
$\ul{g}_{\mu\nu}$. For this application it will be sufficient to
assume that $\alpha$ is constant.

As we have seen the two dark matter fluids are described by the
conserved currents $j^\mu$ and $\ul{j}^\mu$ defined by
Eqs.~\eqref{jmu}. We now suppose that they slightly differ from an
equilibrium configuration described by the equilibrium current
$j_0^\mu=\rho_0 u_0^\mu$, conveniently defined with respect to the
metric $f_{\mu\nu}$, so that $f_{\mu\nu}u_0^\mu u_0^\nu=-1$ and
$\calD_\mu j_0^\mu=0$. In Appendix~\ref{app:linearorder} we give
details of the plasma-like hypothesis. In particular we obtain that
\bes{jjbar} j^\mu &=& j_0^\mu + \calD_\nu\left(j_0^\nu
  y_\perp^\mu-j_0^\mu y_\perp^\nu\right) + \calO\left(2\right)\,,\\
\ul{j}^\mu &=& j_0^\mu + \calD_\nu\left(j_0^\nu
  \ul{y}_\perp^\mu-j_0^\mu \ul{y}_\perp^\nu\right) +
\calO\left(2\right)\,.\ees
Our explicit plasma-like solution is now obtained when we insert the
ansatz~\eqref{jjbar} into the graviphoton field
equation~\eqref{graviphoton}. Indeed, posing for the two displacements
$y^\mu=y_0^\mu+\frac{1}{2}\xi^\mu$ and
$\ul{y}^\mu=y_0^\mu-\frac{1}{2}\xi^\mu$, where
$\xi^\mu=y^\mu-\ul{y}^\mu$ is the relative displacement, we can
straightforwardly integrate this equation with result
\be{Fsol0} W' H^{\mu\nu} = - 4\pi\bigl(j_0^\mu
\,\xi_\perp^\nu-j_0^\nu\,\xi_\perp^\mu\bigr) + \calO\left(2\right)\,.
\ee
This is valid for any function $W(X)$ in the action, where $X$ is
defined by~\eqref{X}, showing that $W' H^{\mu\nu} = \calO(1)$. In the
MOND weak-field regime and also for first-order cosmological
perturbations where $X\ll 1$, the function $W$ behave as
$W'=1+\calO(1)$, see Eq.~\eqref{W0}. Thus Eq.~\eqref{Fsol0} tells us
that $H^{\mu\nu}$ itself is a perturbative quantity, and reduces at
first order to
\be{Fsol} H^{\mu\nu} = - 4\pi\bigl(j_0^\mu
\,\xi_\perp^\nu-j_0^\nu\,\xi_\perp^\mu\bigr) + \calO\left(2\right)\,.
\ee
This solution is analogous to a classic one in relativistic plasma
physics, and is at the basis of our model of dipolar dark matter. It
implies that the stress-energy tensor~\eqref{taumunu} of the internal
field is of second order in the MOND regime and in cosmology:
\be{tauO2} \tau^{\mu\nu} = \calO\left(2\right)\,. \ee
On the other hand, in the limiting case $X\gg 1$ appropriate to the
solar system where we have the postulated behaviour~\eqref{Winf} hence
$W'\sim X^{-b-1}$, Eq.~\eqref{Fsol0} tells us that the dipole moment
scales as $\xi_\perp\sim X^{-b-1/2}$ and can be neglected since
$b>0$. We shall use this result in Sec.~\eqref{sec:1PNlimit} for the
study of the post-Newtonian limit of the theory in the solar system.

We shall now investigate the matter equations and Einstein field
equations at first perturbative order in the weak field limit $X\ll
1$, for which we have already derived the solutions~\eqref{Fsol}
and~\eqref{tauO2}. Inserting~\eqref{Fsol} into the equations of
motion~\eqref{DMeom} of the dark matter particles, and
using~\eqref{umuu0}, we obtain
\bes{DMeompert} a^{\mu} &=& -4\pi \,\alpha^{-3/2}\rho_0
\,\xi_\perp^\mu + \calO\left(2\right)\,,\\ \ul{a}^{\mu} &=& 4\pi
\,\alpha^{3/2}\rho_0 \,\xi_\perp^\mu + \calO\left(2\right)\,. \ees
Thus $a^{\mu}$ and $\ul{a}^{\mu}$ are perturbative quantities of
  order $\calO(1)$.

From now on we shall often view the dark matter, instead of being
composed of the two fluids $j^\mu$ and $\ul{j}^\mu$, as composed of a
single fluid with current $j_0^\mu$, but endowed with the vector field
$\xi_\perp^\mu$. In analogy with the previous model of dipolar dark
matter~\cite{BL08, BL09} we can call the vector field $\xi_\perp^\mu$
a dipole moment. Note that $\xi_\perp^\mu$ is necessarily space-like,
because of the projection orthogonal to the time-like four velocity
$u_0^\mu$ of the equilibrium configuration.

We next make use of the relations~\eqref{jjbar}, or
equivalently~\eqref{rhorho0}--\eqref{umuu0}, to transform the two
equations of motion~\eqref{DMeompert} into two equivalent
equations. First, we obtain the equation of evolution for the dipole
moment,
\be{eqdipole} \ddot{\xi}^\mu_\perp + \xi^\rho_\perp
\mathcal{R}^\mu_{\phantom{\mu}\nu\rho\sigma}u_0^\nu u_0^\sigma =
-\perp^\mu_\sigma\left(2\calD_\nu h^\sigma_\rho-\calD^\sigma
h_{\nu\rho}\right)u_0^\nu u_0^\rho -
4\pi\left(\alpha^{-1/2}+\alpha^{1/2}\right)\rho_0\,\xi_\perp^\mu +
\calO\left(2\right)\,, \ee
where we denote $\ddot{\xi}^\mu_\perp\equiv
u_0^\rho\calD_\rho(u_0^\sigma\calD_\sigma\xi^\mu_\perp)$, and the
Riemann curvature tensor
$\mathcal{R}^\mu_{\phantom{\mu}\nu\rho\sigma}\equiv
R^\mu_{\phantom{\mu}\nu\rho\sigma}[f]$ of the metric $f_{\mu\nu}$
arises from the commutator of covariant derivatives. Second, we get
\be{a0mu0} a_0^\mu + \ddot{y}^\mu_{0\perp} + y^\rho_{0\perp}
\mathcal{R}^\mu_{\phantom{\mu}\nu\rho\sigma}u_0^\nu u_0^\sigma = -
2\pi\left(\alpha^{-1/2}-\alpha^{1/2}\right)\rho_0\,\xi_\perp^\mu +
\calO\left(2\right)\,, \ee
where we pose $\ddot{y}^\mu_{0\perp}\equiv
u_0^\rho\calD_\rho(u_0^\sigma\calD_\sigma y^\mu_{0\perp})$ and recall
that $y_0^\mu=\frac{1}{2}\left(y^\mu+\ul{y}^\mu\right)$. The evolution
of the vector $y_0^\mu$, which is the ``center of position'' of
$y^\mu$ and $\ul{y}^\mu$, is thus governed by~\eqref{a0mu0}. We now
specify the equilibrium configuration by choosing $y_0^\mu=0$, which
implies that the fluid at equilibrium obeys
\be{a0mu} a_0^\mu = -
2\pi\left(\alpha^{-1/2}-\alpha^{1/2}\right)\rho_0\,\xi_\perp^\mu +
\calO\left(2\right)\,. \ee
The equilibrium fluid is geodesic with respect to the metric
$f_{\mu\nu}$ in the special case where the two metrics have the same
background, \textit{i.e.} $\alpha=1$. We shall see that when
the coupling constant $\varepsilon$ is very small (as will be assumed
in Sec.~\ref{sec:NR1PN} to reproduce MOND and to study the 1PN
  limit), $\alpha$ is indeed very close to one so that the equilibrium fluid
is almost geodesic.

For the choice $y_0^\mu=0$ adopted here, we can easily relate the dark
matter stress-energy tensors $T^{\mu\nu}$ and $\ul{T}^{\mu\nu}$ to the
one of the equilibrium fluid, $T_0^{\mu\nu}=\rho_0u_0^\mu u_0^\nu$,
and to the dipole moment $\xi_\perp^{\mu}$ and its time derivative
$\dot{\xi}_\perp^{\mu}\equiv u_0^\rho\calD_\rho\xi_\perp^\mu$:
\bes{TTbar} \beta\,T^{\mu\nu} &=&
\alpha^{-1/2}\biggl[T_0^{\mu\nu}\Bigl(1+\frac{1}{2}h_{\rho\sigma}u_0^\rho
  u_0^\sigma\Bigr) +
  j_0^{(\mu}\dot{\xi}_\perp^{\nu)}-\frac{1}{2}\calD_\rho
  \bigl(\xi_\perp^\rho T_0^{\mu\nu}\bigr)\biggr] +
\calO\left(2\right)\,,\\ \ul{\beta}\,\ul{T}^{\mu\nu} &=&
\alpha^{1/2}\biggl[T_0^{\mu\nu}\Bigl(1-\frac{1}{2}h_{\rho\sigma}u_0^\rho
  u_0^\sigma\Bigr) -
  j_0^{(\mu}\dot{\xi}_\perp^{\nu)}+\frac{1}{2}\calD_\rho
  \bigl(\xi_\perp^\rho T_0^{\mu\nu}\bigr)\biggr] +
\calO\left(2\right)\,. \ees
Concerning the baryons (defined with respect to the metric
$g_{\mu\nu}$) we get the simpler relation
\be{Tbaryon} \beta\,T_\text{b}^{\mu\nu} = \alpha^{-1/2}
\,T_{0\text{b}}^{\mu\nu}\Bigl(1+\frac{1}{2}h_{\rho\sigma}u_{0\text{b}}^\rho
u_{0\text{b}}^\sigma\Bigr) + \calO\left(2\right)\,.\ee

Finally we provide the two Einstein field equations~\eqref{EE12} at
first order in both the metric perturbation and the dipole moment and
in the weak field regime for which we have ${H^{\mu\nu}=\calO(1)}$ and
$\tau^{\mu\nu}=\calO(2)$, according to~\eqref{Fsol}--\eqref{tauO2}. We
apply a standard perturbation analysis to relate both Einstein tensors
$E^{\mu\nu}$ and $\ul{E}^{\mu\nu}$ to the Einstein tensor
$\mathcal{E}^{\mu\nu}$ of the metric $f_{\mu\nu}$ at first order in
the metric perturbation $h_{\mu\nu}$. At zero-th order
$\alpha^{-1}g_{\mu\nu}$ and $\alpha \ul{g}_{\mu\nu}$ reduce to
the same background $f_{\mu\nu}$ and we get a consistency condition on
the matter tensors $T_0^{\mu\nu}$ and $T_{0\text{b}}^{\mu\nu}$ in
Eqs.~\eqref{TTbar}--\eqref{Tbaryon} so that the two corresponding
Einstein field equations for the background are the same:
\be{consistcond} T_{0\text{b}}^{\mu\nu} =
\frac{(\alpha-1)(\varepsilon-1)}{\alpha+\varepsilon} \,T_0^{\mu\nu}
\,.\ee
We thus see that when the two metrics $g_{\mu\nu}$ and
$\ul{g}_{\mu\nu}$ have the same background (\textit{i.e.} $\alpha=1$)
the baryons must be perturbative. In the application to cosmology in
Sec.~\ref{sec:cosmo} we shall adjust the parameter $\alpha$ so
that~\eqref{consistcond} reflects the correct baryonic and dark matter
content of the cosmological background. In addition we find some
constraint relating the constants $\lambda$, $\ul{\lambda}$,
$\lambda_f$ in the original action~\eqref{action}, for the two
backgrounds to be consistent. We shall further restrict this
constraint by requiring that the observed cosmological constant
$\Lambda$ be a true constant even at the level of cosmological
perturbations (see Sec.~\ref{sec:cosmo}). This entails
\be{constcosmo} \lambda = \Lambda\,,\qquad\ul{\lambda} = \alpha^2
\Lambda\,,\qquad\lambda_f = \alpha\,\Lambda \,.\ee

To work out the field equations to first order in perturbations, we
need to control to first order the tensorial coefficients
$\mathcal{A}_{\,\mu\nu}^{\,\rho\sigma}$ and
$\ul{\mathcal{A}}_{\,\mu\nu}^{\,\rho\sigma}$ defined in
Eq.~\eqref{deltaf}. The results are derived in
Appendix~\ref{app:solpert} where we obtain
\be{AApert1} \mathcal{A}_{\,\mu\nu}^{\,\rho\sigma} =
  \frac{1}{\alpha}\left[\delta_{(\mu}^\rho \,\delta_{\nu)}^\sigma -
    h_{(\mu}^{(\rho} \,\delta_{\nu)}^{\sigma)}\right] +
  \calO\left(2\right)\,,\qquad
  \ul{\mathcal{A}}_{\,\mu\nu}^{\,\rho\sigma} =
  \alpha\left[\delta_{(\mu}^\rho \,\delta_{\nu)}^\sigma +
    h_{(\mu}^{(\rho} \,\delta_{\nu)}^{\sigma)}\right] +
  \calO\left(2\right)\,.\ee
As the last ingredient we need also to find the link between the two
Einstein tensors $E^{\mu\nu}$ and $\ul{E}^{\mu\nu}$ and the one
$\mathcal{E}^{\mu\nu}$ of the metric $f_{\mu\nu}$. This is provided by
\bes{EEbar} \beta\,E^{\mu\nu} &=& \mathcal{E}^{\mu\nu}-\frac{1}{2}
\,\Box_\text{L} h^{\mu\nu} +
\calO\left(2\right)\,,\\ \ul{\beta}\,
\ul{E}^{\mu\nu} &=& \mathcal{E}^{\mu\nu}+\frac{1}{2}
\,\Box_\text{L} h^{\mu\nu} + \calO\left(2\right)\,, \ees
where $\Box_\text{L}$ denotes a standard linear operator acting on the
metric perturbation for any background metric
$f_{\mu\nu}$.\,\footnote{Its explicit expression will not be used
	because we only need that $\mathcal{E}^{\mu\nu} =
    \tfrac{1}{2}(\beta E^{\mu\nu} + \ul{\beta} \ul{E}^{\mu\nu}) +
    \calO(2)$, but is given here for completeness:
$$\Box_\text{L} h^{\mu\nu} = \Box \hat{h}^{\mu\nu} -2
    \calD^{(\mu}\hat{H}^{\nu)}+f^{\mu\nu} \calD_{\rho}\hat{H}^{\rho}
    -2\mathcal{C}^{\mu\rho\sigma\nu}\hat{h}_{\rho\sigma}
    -\frac{2}{3}\Bigl(\hat{h}^{\mu\nu}
    -\frac{1}{4}\hat{h}\,f^{\mu\nu}\Bigr)\mathcal{R}\,,$$
where $\Box=\calD_\rho\calD^\rho$,
$\hat{h}^{\mu\nu}=h^{\mu\nu}-\frac{1}{2}f^{\mu\nu}h$,
$\hat{h}=f^{\mu\nu}\hat{h}_{\mu\nu}=-h$,
$\hat{H}^\mu=\calD_{\nu}\hat{h}^{\mu\nu}$, and
$\mathcal{C}^{\mu\rho\sigma\nu}$ and $\mathcal{R}$ denote the Weyl
curvature and scalar curvature of the background.} Finally, we find
that both Einstein field equations can be written into the ordinary
forms
\begin{subequations}\label{EFE1}
\begin{align} E^{\mu\nu} + \Lambda \,g^{\mu\nu} &=
\frac{16\pi}{1+\alpha^2+2\alpha\varepsilon}
\biggl[\alpha(\alpha+2\varepsilon)\bigl(T_\text{b}^{\mu\nu} +
  T^{\mu\nu}\bigr) - \frac{1}{\alpha^4}(1-h)\,\ul{T}^{\mu\nu}
  \nonumber\\ & \qquad\qquad\qquad\quad +
  \dfrac{2}{\alpha^{3/2}}\,h_\rho^{(\mu}\,T_0^{\nu)\rho}\biggr] +
\calO\left(2\right)\,,\label{EFEord} \\[5pt] \ul{E}^{\mu\nu} +
\alpha^2\Lambda \,\ul{g}^{\mu\nu} &= -
\frac{16\pi\,\alpha^2}{1+\alpha^2+2\alpha\varepsilon} \biggl[
  \alpha^4(1+h)\bigl(T_\text{b}^{\mu\nu} + T^{\mu\nu}\bigr) -
  \frac{1+2\alpha\varepsilon}{\alpha^2}\,\ul{T}^{\mu\nu} \nonumber\\ &
  \qquad\qquad\qquad\quad + 2\alpha^{3/2}\frac{1 +
    \alpha\varepsilon}{\alpha+\varepsilon}\,
  h_\rho^{(\mu}\,T_0^{\nu)\rho}\biggr] +
\calO\left(2\right)\,.\label{EFEbar}\end{align}
\end{subequations}
When deriving Eqs.~\eqref{EFE1} we have used the consistency
relation~\eqref{consistcond} and explicitly assumed that $\alpha$ is
constant (if not, further terms have to be added to these equations).


\section{First order cosmological perturbations}
\label{sec:cosmo}

We expand the model around a homogeneous and isotropic
Friedmann-Lema\^{i}tre-Robertson-Walker (FLRW) cosmology, writing both
metrics $g_{\mu\nu}$ and $\ul{g}_{\mu\nu}$ (and therefore also
$f_{\mu\nu}$) as first-order perturbations around some FLRW background
metrics, and solving Eqs.~\eqref{EFE1} by applying cosmological
perturbation techniques to the three metrics. In the end we
shall compare the results with those of the $\Lambda$-CDM model by
looking at the ordinary sector with metric $g_{\mu\nu}$. The other
sector with metric $\ul{g}_{\mu\nu}$ will in principle be unobservable
directly, but since the two sectors are coupled together in the
action~\eqref{action} \textit{via} terms involving the metric
$f_{\mu\nu}$, our solution for the perturbations of the ordinary
sector $g_{\mu\nu}$ will be strongly affected by our solution for the
dark sector $\ul{g}_{\mu\nu}$ and \textit{vice versa}.

\subsection{Background cosmology}
\label{sec:background}

The two background FLRW metric intervals for the two metrics
$g_{\mu\nu}$ and $\ul{g}_{\mu\nu}$ read (with the symbol $\fond{}$
referring to quantities defined in the background):
\bes{FLRW} \fond{\ud s}^{2} &=& a^2\left[- \ud \eta^2 +
  \gamma_{ij}\,\ud x^{i}\,\ud x^{j}\right]\,, \\ \fond{\ud \ul{s}}^{2}
&=& \ul{a}^2\left[- \ud \eta^2 + \gamma_{ij}\,\ud x^{i}\,\ud
  x^{j}\right]\,,\ees
where $\eta$ denotes the conformal time and $a(\eta)$ and
$\ul{a}(\eta)$ are the scale factors, such that $\ud t = a\ud\eta$ and
$\ud \ul{t} = \ul{a}\ud\eta$ are the cosmic time intervals, and $x^i$
are the spatial coordinates. The spatial metric $\gamma_{ij}$, assumed
to be the same for the two backgrounds, is the metric of maximally
symmetric spatial hypersurfaces of constant curvature $K=0$ or $K=\pm
1$. The covariant derivative associated with the spatial metric
$\gamma_{ij}$ will be denoted $\uD_i$. The prime will stand for the
derivative with respect to the conformal time $\eta$, and
$\mathcal{H}\equiv a'/a$ and $\ul{\mathcal{H}}\equiv \ul{a}'/\ul{a}$
denote the conformal Hubble parameters. Solving~\eqref{fdef} we obtain
the FLRW background for the metric $f_{\mu\nu}$,
\be{FLRWf} \fond{\ud s}_{\!f}^{2} = a\ul{a}\left[- \ud \eta^2 +
  \gamma_{ij}\,\ud x^{i}\,\ud x^{j}\right]\,, \ee
whose scale factor is $\sqrt{a\ul{a}}$. Recall that we
introduced the parameter $\alpha$ in our perturbation
assumptions~\eqref{ggbar} to account for the fact that the baryons
have been inserted in the ordinary sector with metric $g_{\mu\nu}$ but
not in the dark sector with metric $\ul{g}_{\mu\nu}$. We thus see
that, in cosmology,
\be{alpha} \alpha = \frac{a}{\ul{a}}\,. \ee
Since $\alpha$ has been assumed from the start in Sec.~\ref{sec:displ}
to be constant we are thus looking for two background cosmologies with
identical Hubble parameters,\,\footnote{Note that this also agrees
  with $\mathcal{H}_f = \frac{1}{2}(\mathcal{H} + \ul{\mathcal{H}})$.}
\be{HHbar} \mathcal{H} = \ul{\mathcal{H}}\,. \ee
We also assume that the three matter fluids are comoving in their
respective backgrounds, hence their background velocities read
\be{fonduubar} \fond{u}_\text{b}^\mu = \,\fond{u}^\mu =
\Bigl(\frac{1}{a}, \mathbf{0}\Bigr)\,,\qquad \fond{\ul{u}}^\mu =
\Bigl(\frac{1}{\ul{a}}, \mathbf{0}\Bigr)\,.\ee
The background matter densities obey the standard evolution laws
\be{evolulaw} \fond{\rho}_\text{b}' +
3\mathcal{H}\fond{\rho}_\text{b} = 0\,,\qquad \fond{\rho}' +
3\mathcal{H}\fond{\rho} = 0\,,\qquad \fond{\ul{\rho}}' +
3\mathcal{H}\fond{\ul{\rho}} = 0\,.\ee
In Sec.~\ref{sec:displ} we have shown how the two dark matter fluids
$\rho$, $u^\mu$ and $\ul{\rho}$, $\ul{u}^\mu$ are related together
through the equilibrium fluid configuration $\rho_0$, $u_0^\mu$, see
Eqs.~\eqref{jjbar} or equivalently~\eqref{rhorho0}--\eqref{umuu0}. In
particular, such relations imply that in the background the two dark
matter fluid densities obey 
\be{solevolulaw} \fond{\rho} =
\alpha^{-3/2}\fond{\rho}_0\,,\qquad\fond{\ul{\rho}} =
\alpha^{3/2}\fond{\rho}_0\,, \ee
which implies that $\fond{\ul{\rho}} = \alpha^3 \fond{\rho}$. Hence
Eqs.~\eqref{evolulaw} can be solved as
\be{solevolulawbar} \fond{\rho}_\text{b} =
\frac{k_\text{b}}{a^3}\,,\qquad\fond{\rho} =
\frac{k}{a^3}\,,\qquad\fond{\ul{\rho}} = \frac{k}{\ul{a}^3}\,, \ee
with $k_\text{b}$ and $k$ denoting two constants. Note also that the
equilibrium fluid $\rho_0$, $u_0^\mu$ is obviously given in the
background by
\be{backgrequil} \fond{u}_0^\mu = \Bigl(\frac{1}{(a\ul{a})^{1/2}},
\mathbf{0}\Bigr)\,,\qquad\fond{\rho}_0 = \frac{k}{(a\ul{a})^{3/2}}\,.
\ee

The Friedmann equations of the two backgrounds are now obtained from
Eqs.~\eqref{EFE1} or, alternatively, directly from Eqs.~\eqref{EE12}
as
\bes{Friedmann} 3\bigl(\mathcal{H}^2 + K\bigr) - \Lambda\,a^2 &=&
\frac{16\pi}{1+\alpha^2+2\alpha\varepsilon}\Bigl[\alpha\left(\alpha+2\varepsilon\right)\left(\fond{\rho}_\text{b}
  +\fond{\rho}\right)-\frac{1}{\alpha^2}\fond{\ul{\rho}}\Bigr]a^2
\,,\label{Friedmannord} \\ 3\bigl(\ul{\mathcal{H}}^2 + K\bigr) -
\alpha^2\Lambda\,\ul{a}^2 &=&
\frac{16\pi}{1+\alpha^2+2\alpha\varepsilon}\Bigl[-\alpha^4\left(\fond{\rho}_\text{b}
  +\fond{\rho}\right)+\left(1+
  2\alpha\varepsilon\right)\fond{\ul{\rho}}\Bigr]\,\ul{a}^2\,.\ees
Finally we must impose the equivalence between the two Friedmann
equations~\eqref{Friedmann}. The left-hand sides of these equations
are obviously consistent because $\mathcal{H} = \ul{\mathcal{H}}$ and
$\alpha=a/\ul{a}$. Now the consistency of the right-hand sides is
ensured by the condition
\be{kkb} k_\text{b} =
\frac{(\alpha-1)(\varepsilon-1)}{\alpha+\varepsilon} \,k\,,\ee
which is nothing but the general relation~\eqref{consistcond} when
translated to the case of comoving fluids in a FLRW
background. Physically it states how the ratio between the two scale
factors $\alpha=a/\ul{a}$ is to be related to the relative proportion
of baryonic and dark matter in the two cosmological backgrounds, given
that the baryons have been included into the ordinary sector of the
action~\eqref{action} but not into the dark sector (nor in the
interacting sector). Thus, with this condition, the total matter
density seen in the background of the ordinary sector (and thus
directly measurable in cosmology) reads
\be{rhoM} \fond{\rho}_\text{M} =
\frac{2\alpha\varepsilon}{\alpha+\varepsilon}\fond{\rho}\,.\ee
When studying cosmological perturbations it will be convenient to
define \textit{separately} the effective baryonic and dark matter
densities as seen in the ordinary sector:
\be{effdensity} \fond{\rho}_\text{B} =
\frac{2\alpha(\alpha+2\varepsilon)}{1+\alpha^2+2\alpha\varepsilon}
\fond{\rho}_\text{b}\,,\qquad\fond{\rho}_\text{DM} =
\frac{2\alpha(\alpha-1+2\varepsilon)}{1+\alpha^2+2\alpha\varepsilon}
\fond{\rho}\,.\ee
These definitions come directly from the right-hand side of the
Friedmann equation~\eqref{Friedmannord} in the ordinary sector
and satisfy $\fond{\rho}_\text{M} =
\,\fond{\rho}_\text{B}+\fond{\rho}_\text{DM}$. Let us then suppose
that there is a fraction $p$ of baryons with respect to the total
matter, so that
\be{fraction} \frac{\fond{\rho}_\text{B}}{\fond{\rho}_\text{M}} =
\frac{1}{p}\,.\ee
According to the latest results from Planck we have $p \simeq
6.4$~\cite{Planck}. Computing the ratio~\eqref{fraction} from
Eqs.~\eqref{rhoM} and~\eqref{effdensity} and solving for $\alpha$, we
obtain an analytic expression in terms of the baryonic fraction $p$
and the coupling constant $\varepsilon$.\,\footnote{It reads
  explicitly
$$\alpha(p,\varepsilon) =
  \frac{p(1-3\varepsilon+2\varepsilon^2)-2\varepsilon^2 +
    \sqrt{(1-\varepsilon)\bigl[p^2(1+3\varepsilon-4\varepsilon^3)+4p
        \varepsilon (-1+\varepsilon+2\varepsilon^2)
        -4\varepsilon^2(1+\varepsilon)\bigr]}}
       {2\bigl[p(1-\varepsilon)+\varepsilon\bigr]}\,.$$} 

In Sec.~\ref{sec:NR1PN} we shall recover the MOND phenomenology for
dark matter in galaxies and the correct post-Newtonian limit in the
solar system when $\varepsilon\ll 1$. The interesting application of
the present model will therefore be the limit where $\varepsilon\to
0$, in which case we get
\be{limitalpha} \alpha = 1 - \frac{2\varepsilon}{p} +
\calO\left(\varepsilon^3\right)\,.\ee
Our conclusion is that, although we shall work out the cosmology of
the model for an arbitrary parameter $\alpha$ and a general coupling
constant $\varepsilon$, we can always have in mind that $\alpha$ is
very close to one, hence the two backgrounds of $g_{\mu\nu}$ and
$\ul{g}_{\mu\nu}$ are very close to each other. This means in
particular that the equilibrium dark matter fluid is almost geodesic
with respect to the metric $f_{\mu\nu}$. Indeed
$a_0^\mu=\calO(\varepsilon)$ from Eq.~\eqref{a0mu}, which constitutes
a useful fact further discussed in Sec.~\ref{sec:1PNlimit}. Note also
that Eq.~\eqref{rhoM} tells us that in the limit $\varepsilon\to 0$,
the measured matter density at cosmological scales is
\be{rhoMestimate} \fond{\rho}_\text{M} \sim 2\varepsilon\!\fond{\rho}
\sim 10^{-29}\,\mathrm{g}\,\mathrm{cm}^{-3}\,,\ee
which is much smaller than the ``bare'' dark matter density
$\fond{\rho}$ which has been introduced into the action~\eqref{action}
and could take a huge value. By extension we see that in the
limit $\varepsilon\to 0$, the density of baryons $\rho_\text{b}$
should be much smaller than the ``bare'' density of dark matter $\rho$
in the initial action~\eqref{action}. The baryons could be seen as
resulting from a small ``symmetry breaking'' between the ordinary and
dark sectors of the model.

\subsection{First order perturbations in the ordinary sector}
\label{sec:pertEFE}

We already concluded in Sec.~\ref{sec:background} that the background
evolution is standard, driven by a cosmological constant and by the
matter density defined by~\eqref{rhoM}. We shall now show that the
perturbation equations for the metric $g_{\mu\nu}$, which in our model
represents the metric felt by the baryons and ordinary matter fields
(including ordinary electromagnetic radiation), take the same form as
those for the $\Lambda$-CDM model.\,\footnote{We have imposed the
  relations~\eqref{constcosmo} in order to have a true cosmological
  constant in the background and at the level of perturbations. We
  could have imposed weaker conditions such that it would be constant
  only in the background, but would deviate from a pure cosmological
  constant at the first order in perturbations.} For the sake of
clarity we relegate the definition of standard gravitational and
matter perturbations to Appendix~\ref{app:cosmopert}.

We now introduce new effective variables describing the
dark matter seen in first order cosmological perturbations. In terms
of these variables the perturbation equations for the ordinary metric
$g_{\mu\nu}$ in our model take the standard form. The effective
density contrast and SVT velocity of dark matter are defined by
\bes{effDM} \delta^\text{F}_\text{DM} &=& \delta^\text{F} -
\frac{\Delta z - dA}{\alpha-1+2\varepsilon}\,,\\V_\text{DM} &=& V +
\frac{z' + \frac{1}{2}\,dB}{\alpha-1+2\varepsilon}\,,\\ V^i_\text{DM}
&=& V^i + \frac{{z'}^i +
  \frac{1}{2}\,dB^i}{\alpha-1+2\varepsilon}\,,\ees
together with the usual variables $\delta^\text{F}_\text{b}$,
$V_\text{b}$ and $V^i_\text{b}$ for the baryons. All relevant
  quantities are introduced in Appendix~\ref{app:cosmopert}, notably
  the dipole moment variables $z$ and $z^i$ defined in~\eqref{zzi}.
Furthermore we shall use the effective background baryonic and dark
matter densities defined by Eqs.~\eqref{effdensity}. With those
definitions we find the following gravitational perturbation equations
for the scalar, vectorial and tensorial modes in the \textit{ordinary}
sector $g_{\mu\nu}$
\bes{gravperteq} &&\Delta\Psi - 3\mathcal{H}^2 X = 4 \pi
\,a^2\Bigl(\fond{\rho}_\text{B}\,\delta^\text{F}_\text{b} +
\fond{\rho}_\text{DM}\,
\delta^\text{F}_\text{DM}\Bigr)\,,\\ &&\Psi-\Phi = 0\,,\\ &&\Psi' +
\mathcal{H} \Phi = -4\pi \,a^2 \Bigl(\fond{\rho}_\text{B}\,V_\text{b}
+ \fond{\rho}_\text{DM}\,V_\text{DM}\Bigr)\,,\\ &&\mathcal{H}
X'+(\mathcal{H}^2+2\mathcal{H}')X = 0\,,\label{eqX}\\ &&(\Delta+
2K)\Phi^i = - 16\pi \,a^2 \Bigl(\fond{\rho}_\text{B}\,V^i_\text{b} +
\fond{\rho}_\text{DM}\,V^i_\text{DM}\Bigr)\,,\\ &&{\Phi'}^i + 2
\mathcal{H}\Phi^i = 0\,,\\ &&{E''}^{ij}+2\mathcal{H}{E'}^{ij}+(2K -
\Delta)E^{ij} = 0\,,\label{eqEij}\ees
where the unknowns are the five gravitational variables $\Psi$,
$\Phi$, $X$, $\Phi^i$, $E^{ij}$ and the six matter variables
$\delta^\text{F}_\text{DM}$, $V_\text{DM}$, $V^i_\text{DM}$ and
$\delta^\text{F}_\text{b}$, $V_\text{b}$, $V^i_\text{b}$. Recall that
according to Eq.~\eqref{Xdef}, $X$ is not independent from the other
variables. 

As the equations~\eqref{gravperteq} are exactly the same as the
perturbation equations of the standard cosmological
model~\cite{PUcosmo}, we conclude that the present model is
indistinguishable from standard $\Lambda$-CDM at the level of first
order perturbations, and therefore should reproduce the observed
anisotropies of the CMB.
Indeed, these equations can be evolved without any reference to the
dipole moment, which is unobservable in cosmology (but which will play
a crucial role at galactic scales, see Sec.~\ref{sec:NRlimit}). Note
also that this result is obtained for any value of the coupling
constant $\varepsilon$, as this coupling constant has been absorbed
into the definition of the effective matter
densities~\eqref{rhoM}--\eqref{effdensity}, and that the MOND
acceleration scale $a_0$ does not appear at this level in cosmology.

To be consistent with the field equations~\eqref{gravperteq} and with
the equations of motion for the baryons which are standard, the
effective dark matter variables introduced in Eqs.~\eqref{effDM} must
obey the continuity equation
\be{contDM} {\delta'}^\text{F}_\text{DM} + \Delta V_\text{DM} =
0\,,\ee
together with the Euler equations
\bes{EOMDM} V'_\text{DM} + \mathcal{H} V_\text{DM} + \Psi &=&
0\,,\\ {V'}^i_\text{DM} + \mathcal{H} V^i_\text{DM} &=& 0\,.\ees
The standard form of Eqs.~\eqref{contDM}--\eqref{EOMDM} means that the
effective dark matter described by the effective
variables~\eqref{effDM} obeys the ordinary geodesic equation with
respect to the metric $g_{\mu\nu}$.
In principle, all other variables in the model are unobservable using
current cosmological observations performed in the ordinary sector.

Besides the ordinary sector we have similar equations for the dark
sector $\ul{g}_{\mu\nu}$. It is very important to check that the
latter equations are consistent with Eqs.~\eqref{gravperteq} and
permit to determine all the variables of the model, even those that
are unobservable in the ordinary sector. The full investigation of the
dark sector is relegated to the Appendix~\ref{app:pertEFEother} where
we shall see that the equations~\eqref{contDM}--\eqref{EOMDM} can
equivalently be obtained from the perturbation equations in the dark
sector $\ul{g}_{\mu\nu}$. In particular the continuity and Euler
equations~\eqref{contDM}--\eqref{EOMDM} are consistent with the
equations of motion~\eqref{conteq} and~\eqref{eomSVT}, provided that
the equations in the dark sector are satisfied. Finally, we show in
Appendix~\ref{app:pertEFEother} that all variables in the model can be
determined by solving well-defined linear evolution equations.


\section{Non relativistic and post-Newtonian limits}
\label{sec:NR1PN}

\subsection{Phenomenology of MOND at galactic scales}
\label{sec:NRlimit}

In this section, we investigate the non-relativistic (NR) limit of our
model (\textit{i.e.} formally when the speed of light $c\to +\infty$)
and recover the Bekenstein \& Milgrom~\cite{BekM84} modified Poisson
equation for the gravitational field. The MOND function $\mu$ that we
shall obtain is directly related to the function $W$ introduced into
the action~\eqref{action}. We have already adjusted this function in
Eqs.~\eqref{W0}--\eqref{Winf} in such a way that the model will be in
agreement with the phenomenology of MOND at galactic
scales~\cite{Milg1, Milg2, Milg3}. Furthermore, thanks to this
adjustment we shall investigate the model in the solar system in
Sec.~\ref{sec:1PNlimit}.

We now work out the NR limit directly at the level of the
action~\eqref{action}. For convenience we restore for a while the
gravitational constant $G$ and the speed of light $c$ such that the
action has the dimension of the Planck constant. We insert into the
action the standard ansatz for the metric at lowest order, namely
\be{Udef} g_{00} = -1 + \frac{2 \,U}{c^2} +
\calO\left(c^{-4}\right)\,,\ee
together with $g_{0i} = \calO(c^{-3})$ and $g_{ij} = \delta_{ij} +
\calO(c^{-2})$, where $U$ represents the ordinary Newtonian potential
felt by ordinary baryonic matter and $\calO(c^{-n})$ denotes the small
post-Newtonian remainder. Similarly we write\,\footnote{Thus the two
  metrics $g_{\mu\nu}$ and $\ul{g}_{\mu\nu}$ (and $f_{\mu\nu}$ as
  well), differ by small post-Newtonian corrections from the same
  Minkowskian background, which implies $\alpha=1$ in the notation of
  Eqs.~\eqref{ggbar}. We adopt $\alpha=1$ for this application,
  all-over the present section and also in the next
  one~\ref{sec:1PNlimit}.}
\be{Ubardef} \ul{g}_{00} = -1 + \frac{2 \,\ul{U}}{c^2} +
\calO\left(c^{-4}\right)\,,\ee
and $\ul{g}_{0i} = \calO(c^{-3})$, $\ul{g}_{ij} = \delta_{ij} +
\calO(c^{-2})$, where $\ul{U}$ is the Newtonian potential of the dark
sector. We also write a similar ansatz for the vector field $K_\mu$,
namely
\be{phidef} K_{0} = \frac{\phi}{c^2} + \calO\left(c^{-4}\right)\,,\ee
with $K_{i} = \calO(c^{-3})$, where $\phi$ denotes an appropriate
Coulomb-type potential. For the dipole vector field our ansatz
is
\be{xidef} \xi_\perp^i = \lambda^i + \calO(c^{-2})\,,\ee
 where $\lambda^i$ is the dipole moment in the NR limit, together with
 $\xi_\perp^0 = \calO(c^{-1})$ which is consistent with
 $u_{0\mu}\xi_\perp^\mu=0$ in the NR limit.

The baryonic and dark matter particles are described by their
Newtonian coordinate densities $\rho^*_\text{b}$, $\rho^*$ and
$\ul{\rho}^*$ and their Newtonian coordinate velocities
$\bm{v}_\text{b}$, $\bm{v}$ and $\ul{\bm{v}}$.\,\footnote{We use
  boldface notation to represent ordinary three-dimensional Euclidean
  vectors.} These quantities are linked by the usual continuity
equations, for instance
$\partial_t\rho^*+\bm{\nabla}\cdot(\rho^*\bm{v}) = 0$. It is well
known that the NR limit has to be performed holding these variables
fixed. Furthermore, denoting by $\bm{v}_0$ the ordinary velocity of
the equilibrium configuration we get from Eqs.~\eqref{umuu0}
$\bm{v}=\bm{v}_0+\frac{1}{2}\frac{\ud\bm{\lambda}}{\ud
  t}+\calO(c^{-2})$ and
$\ul{\bm{v}}=\bm{v}_0-\frac{1}{2}\frac{\ud\bm{\lambda}}{\ud
  t}+\calO(c^{-2})$, where $\frac{\ud}{\ud t} =
\partial_t+\bm{v}_0\cdot\nabla$ is the usual convective time
derivative. Thus
\be{vivibar} \bm{v}-\ul{\bm{v}} = \frac{\ud\bm{\lambda}}{\ud
    t}+\calO(c^{-2})\,.\ee
Note also that
  $\xi_\perp^0=\frac{1}{c}\bm{v}\cdot\bm{\lambda}+\calO(c^{-3})$.

The non-relativistic action $S_\text{NR}$ is defined as the limit when
$c\to +\infty$ of the action $S$ to which we substract the
contributions coming from the rest masses of the particles, for
instance $m^*=\int\ud^3\mathbf{x}\,\rho^*$, namely
\be{limitNR} S_\text{NR} = \lim_{c\to +\infty} \left[S
  +\bigl(m_\text{b}^*+m^*+\ul{m}^*\bigr)c^2\int \ud t\right]\,.\ee
The NR limit is straightforwardly computed from the
action~\eqref{action} using the fact that the Ricci scalar density
admits the limit $\sqrt{-g}R =-\frac{2}{c^4}\bigl(\bm{\nabla}U\bigr)^2
+\text{div} +\calO(c^{-6})$ where we can discard the total divergence
which does not contribute to the dynamics. We obtain
\begin{eqnarray}
S_\text{NR} &=& \int\ud t\,\ud^{3}\mathbf{x}\left\{ -\frac{1}{16\pi
  G}\Bigl[\bigl(\bm{\nabla}U\bigr)^2 + \bigl(\bm{\nabla}\ul{U}\bigr)^2
  +\frac{1}{2\varepsilon}\bigl(\bm{\nabla}[U+\ul{U}]\bigr)^2 - 2a_0^2
  \,W\left(X\right)\Bigr]\right.\nonumber\\&&
\qquad\qquad\quad\left. +
\rho_\text{b}^*\Bigl(U+\frac{\bm{v}^2_\text{b}}{2}\Bigr) +
\rho^*\Bigl(U+\phi+\frac{\bm{v}^2}{2}\Bigr)+
\ul{\rho}^*\Bigl(\ul{U}-\phi+\frac{\ul{\bm{v}}^2}{2}\Bigr)\right\}\,,
\label{actionNR}
\end{eqnarray}
where $X=(\bm{\nabla}\phi)^2/a_0^2$ in the NR limit. Note that when
applying the NR limit we assume that the cosmological constant
parameters $\lambda$, $\ul{\lambda}$, $\lambda_f$ scale like
$\Lambda\sim a_0^2/c^4$ and are therefore negligible when $c\to\infty$
(see Ref.~\cite{BL08} for a discussion). The NR
action~\eqref{actionNR} is independent of $c$ and from now on we
conveniently redefine $G=1$.

We then vary the action with respect to all fields and particles. Of
course, the results can alternatively be obtained as the NR limit of
the relativistic equations derived in Sec.~\ref{sec:Model}. The
baryons obey the standard Newtonian law of dynamics,
\be{eomNRb} \frac{\ud \bm{v}_\text{b}}{\ud t} = \bm{\nabla} U\,,\ee
but because of the internal potential $\phi$, the dark matter
particles receive a supplementary Coulomb-type acceleration,
\bes{eomNR} \frac{\ud \bm{v}}{\ud t} &=&
\bm{\nabla}\bigl(U+\phi\bigr)\,,\\\frac{\ud \ul{\bm{v}}}{\ud t} &=&
\bm{\nabla}\bigl(\ul{U}-\phi\bigr)\,,\ees
where the Coulomb potential $\phi$ obeys the modified Gauss
equation
\be{eqgauss} \bm{\nabla}\cdot\Bigl[W'\,\bm{\nabla}\phi \Bigr] =
4\pi\left(\rho^* - \ul{\rho}^*\right)\,,\ee
and we recall that $W'=\ud W/\ud X$. Note that Eqs.~\eqref{eomNR}
imply that $\ud\bm{v}_0/\ud t=\frac{1}{2}\bm{\nabla}(U+\ul{U})$ which
is consistent with $a_0^\mu=0$, as we have found in Eq.~\eqref{a0mu}
with $\alpha=1$. Finally, the Newtonian potentials $U$ and $\ul{U}$
obey two equations, which can be re-arranged into
\bes{eqUUbar} &&\Delta U = - \frac{4\pi}{1+\varepsilon}
\Bigl[\bigl(1+2\varepsilon\bigr) \bigl(\rho_\text{b}^*+\rho^*\bigr)
  -\ul{\rho}^*\Bigr]\,,\label{eqUUbara}\\ &&
\Delta\bigl(U+\ul{U}\bigr) = - \frac{8\pi\varepsilon}{1+\varepsilon}
\bigl(\rho_\text{b}^*+\rho^*+\ul{\rho}^*\bigr)\,.\label{eqUUbarb}\ees

With these equations in hands we now look for a plasma-like
solution. Namely, the densities $\rho^*$ and $\ul{\rho}^*$ are related
to the density $\rho_0^*$ of the equilibrium configuration by
\bes{rho*sol} \rho^* &=& \rho_0^* -
\frac{1}{2}\,\bm{\nabla}\cdot\bm{P}\,,\\ \ul{\rho}^* &=& \rho_0^* +
\frac{1}{2}\,\bm{\nabla}\cdot\bm{P}\,.\ees
In these relations, which represent the NR limit of
Eqs.~\eqref{rhorho0}, we define the polarization field $\bm{P} =
\rho_0^*\,\bm{\lambda}$, with $\bm{\lambda}$ being the NR limit of the
dipole moment in
Eq.~\eqref{xidef}. Inserting~\eqref{rho*sol} into~\eqref{eqgauss} and
integrating we obtain
\be{polar} W'\,\bm{\nabla}\phi = - 4\pi\,\bm{P} \,,\ee
which is the NR limit of Eq.~\eqref{Fsol0}. Thus, quite naturally the
internal force field is aligned with the polarization vector.

Let us now show that a mechanism of ``gravitational polarization''
takes place when the coupling constant $\varepsilon$ is very small,
$\varepsilon \ll 1$. Indeed, we expect from the form of the coupling
term in~\eqref{actionNR} that the latter condition will enforce the
two potentials $U$ and $\ul{U}$ to be opposite to each other. In the
limit $\varepsilon\ll 1$, Eq.~\eqref{eqUUbarb} reduces to
$\Delta(U+\ul{U}) = 0$, hence we can take $U+\ul{U} = 0$. Then
Eq.~\eqref{eqUUbara} reduces to a simple Poisson equation for the
ordinary Newtonian potential felt by baryonic matter,\,\footnote{See
  the end of Sec.~\ref{sec:1PNlimit} for the discussion of a residual
  dark matter contribution $\rho_\text{DM}^*=2\varepsilon\rho^*$
  coming from the right-side of Eq.~\eqref{eqUUbara}.}
\be{Uord} \Delta U = - 4\pi\bigl(\rho_\text{b}^*+\rho^* -
\ul{\rho}^*\bigr) \,,\ee
while the equations of motion of the dark matter particles now read
\bes{eomNRsol} \frac{\ud \bm{v}}{\ud t} &=&
\bm{\nabla}\bigl(U+\phi\bigr)\,,\\\frac{\ud \ul{\bm{v}}}{\ud t} &=& -
\bm{\nabla}\bigl(U+\phi\bigr)\,.\ees

With this mechanism we observe that the ``effective'' gravitational to
inertial mass ratio $m_\text{g}/m_\text{i}$ of the two species of dark
matter particles is $\pm 1$, and we can interpret the dark matter
medium as a ``gravitational plasma'' composed of particles with masses
$(m_\text{i},m_\text{g})=(m,\pm m)$ interacting \textit{via} the
gravito-electric field $\phi$ generated by the gravitational masses
(or charges) $m_\text{g}=\pm m$ (see~\cite{B07mond, BBwag} for further
discussions). We however note that in the present model no negative
masses have been introduced, since each species of dark matter
particles in the relativistic action~\eqref{action} has been coupled
in a standard way to its respective metric.

In such a gravitational plasma the particles reach equilibrium when
the internal force exactly balances the gravitational field, namely
\be{equil} \bm{\nabla} \phi = - \bm{\nabla} U\,.\ee
At equilibrium the dark matter fluid is unaccelerated (in the ordinary
three-dimensional sense) while the ordinary matter is accelerated in
the standard way. Under this condition the polarization
field~\eqref{polar} at equilibrium is therefore
\be{polareq} \bm{P} = \frac{W'}{4\pi}\,\bm{\nabla} U\,,\ee
where $W'(X)$ is now a function of the norm of the gravitational field
through $X=(\bm{\nabla}U)^2/a_0^2$. At equilibrium the polarization
$\bm{P}$ is thus aligned with the local value of the gravitational
field $\bm{g}=\bm{\nabla}U$, which is what we mean by ``gravitational
polarization''.

Finally the MOND equation follows immediately from Eq.~\eqref{Uord},
which can be transformed thanks to~\eqref{rho*sol} into
\be{Upol} \bm{\nabla}\cdot\Bigl[\bm{\nabla}U - 4\pi \bm{P}\Bigr] = -
4\pi\,\rho_\text{b}^* \,.\ee
Using the constitutive relation~\eqref{polareq} the latter equation
takes exactly the form of the modified Poisson equation~\cite{BekM84}:
\be{BMeq} \bm{\nabla}\cdot\biggl[\mu\left(
  \frac{\vert\bm{\nabla}U\vert}{a_0}\right) \bm{\nabla}U \biggr] = -
4\pi\,\rho_\text{b}^* \,,\ee
where the MOND interpolating function is given by $\mu = 1 - W'$. It
is then easy to see that with the postulated form~\eqref{W0} of the
function $W$ in the regime $X\to 0$, one recovers the correct MOND
regime when $g\ll a_0$, namely
\be{muMOND} \mu = 1 - W' = \frac{g}{a_0} +
\calO\left(\frac{g^2}{a_0^2}\right)\,.\ee

On the other hand, we want to recover the ordinary Poisson equation in
the Newtonian regime $g\gg a_0$. From
Eqs.~\eqref{BMeq}--\eqref{muMOND} we see that it suffices to impose
that $W'(X)$ tends to zero in the formal limit when $X\to
+\infty$. However, in order to suppress any residual
polarization~\eqref{polareq} when $g\gg a_0$, we prefer to impose the
stronger condition that $\sqrt{X}\,W'\to 0$ when $X\to \infty$, hence
the behaviour postulated in Eq.~\eqref{Winf}. The choice $b>0$ rather
than $b>-\frac{1}{2}$ is to ensure that $W$ remains finite in the
limit $X\to \infty$. In the next section~\ref{sec:1PNlimit} we shall
study the first post-Newtonian (1PN) approximation of the theory in
the solar system under the assumption~\eqref{Winf}.

It remains to show that the equilibrium defined by the
condition~\eqref{equil} is stable. To prove it we show that the dark
matter medium undergoes stable plasma-like oscillations. Indeed, by
computing the relative acceleration of the two particle species
combining Eqs.~\eqref{eomNRsol} and~\eqref{vivibar}, and using the
solution~\eqref{polar} for the internal field, we obtain the following
harmonic oscillator governing the evolution of the dipole moment
$\bm{\lambda}$:\,\footnote{This equation can also be recovered from
  the more general equation of evolution of the dipole
  moment~\eqref{eqdipole}.}
\be{harmosc} \frac{\ud^2 \bm{\lambda}}{\ud t^2} + \omega^2
\bm{\lambda} = 2 \bm{\nabla} U\,.\ee
The derivation is of course analogous to the classic derivation of the
plasma oscillations in electrodynamics~\cite{Jackson}. The plasma
frequency we get in the present context reads
\be{plasma} \omega = \sqrt{\frac{8\pi\rho_0^*}{W'}}\,.\ee
In the MOND regime we have $W'\to 1$, and this frequency is simply the
one associated with the self-gravitating dynamical time scale
$\tau=\frac{2\pi}{\omega}=\sqrt{\frac{\pi}{2\rho_0^*}}$.

\subsection{Post-Newtonian limit in the Solar System}
\label{sec:1PNlimit}

In this section we investigate the theory in the regime of the Solar
System (SS) where $g\gg a_0$ hence $X\gg 1$. We have already
postulated in Eq.~\eqref{Winf} the form of the function $W(X)$ in this
regime,
\be{Winf'} W(X)= A + \frac{B}{X^{b}} +
o\!\left(\frac{1}{X^{b}}\right)\,,\ee
in which $b>0$. With this choice we have seen that we recover the
usual Poisson equation~\eqref{BMeq} since $W'\to 0$, and we suppress
any polarization effect in the NR limit since $\sqrt{X}\,W'\to 0$, see
Eq.~\eqref{polareq}. Furthermore it is clear that the suppression of
polarization effects goes beyond the NR limit. Indeed
Eq.~\eqref{Fsol0} tells us that when $\sqrt{X}\,W'\to 0$ the dipole
moment $\xi^\mu_\perp$ is negligible and therefore the dark matter
medium becomes inactive.

In addition we want to impose that $W(X)$ itself tends to zero or a
constant in the limit $X\to +\infty$, which is the reason for our
choice $b>0$. The constant $A$ will simply add to the value of
the cosmological constant in the regime $g\gg a_0$. Our conclusion is
that the action~\eqref{action} in the strong field regime $g\gg
  a_0$ reduces to
\begin{eqnarray} S_\text{strong field} &=& \int\ud^{4}x 
\left\{ \frac{\sqrt{-g}}{32\pi}\bigl(R-2\lambda\bigr) +
\frac{\sqrt{-\ul{g}}}{32\pi}\bigl(\ul{R}-2\ul{\lambda}\bigr) +
\frac{\sqrt{-f}}{16\pi \varepsilon}\bigl(\mathcal{R}-2\lambda'_f\bigr)
\right.\nonumber\\&&\qquad\quad\left. - \sqrt{-g}\,\rho_\text{b} - 2
\sqrt{-f}\,\rho_0\right\} \,,\label{actionSS}\end{eqnarray}
where we have posed $\lambda'_f = \lambda_f -
\varepsilon\,a_0^2\,A$. To derive~\eqref{actionSS} we used the fact
that when $\xi^\mu_\perp$ is negligible the coupling between the
currents $j^\mu$ and $\ul{j}^\mu$ and the graviphoton field $K_\mu$
disappears because $j^\mu=\ul{j}^\mu$ from Eqs.~\eqref{jjbar}. Note
the residual contribution of dark matter in this action, and that we
shall discuss at the end of this section.\,\footnote{Here $\rho_0$ is
  the density of dark matter in the equilibrium configuration defined
  with respect to $f_{\mu\nu}$. Its contribution in~\eqref{actionSS}
  comes from Eqs.~\eqref{rhorho0} in the case $\alpha=1$, and is valid
  only up to second order terms $\calO(2)$, negligible for the present
  discussion.}

Here we shall explore the consequences of the action~\eqref{actionSS}
in a post-Newtonian context, to study the first post-Newtonian (1PN)
limit of this theory in the SS. As usual we can neglect all
cosmological constant terms in the SS. The ordinary metric
$g_{\mu\nu}$ at 1PN order is parametrized by two potentials, the
``gravitoelectric'' scalar potential $V$ and the ``gravitomagnetic''
vector potential $V_i$, say $g^\text{1PN}_{\mu\nu}=g_{\mu\nu}[V,
  V_i]$, by which we mean that
\bes{g1PN} g_{00} &=& -1 + \frac{2 V}{c^2} - \frac{2 V^2}{c^4} +
\calO\left(c^{-6}\right)\,,\\ g_{0i} &=& - \frac{4 V_i}{c^3} +
\calO\left(c^{-5}\right)\,,\\ g_{ij} &=&
\delta_{ij}\,\Bigl(1+\frac{2V}{c^2}\Bigr) +
\calO\left(c^{-4}\right)\,.\ees
In exactly the same way we parametrize the 1PN metric in the dark
sector with two other 1PN potentials $\ul{V}$ and $\ul{V}_i$, namely
$\ul{g}^\text{1PN}_{\mu\nu}=\ul{g}_{\mu\nu}[\ul{V},
  \ul{V}_i]$.\,\footnote{The two forms of the metrics that we
  postulated above will be justified when we find a consistent
  solution of the 1PN equations.} The point now is to find the 1PN
parametrization of the metric $f_{\mu\nu}$ in the interacting sector
of the action~\eqref{actionSS}. For this purpose we make use of the
result derived in Eq.~\eqref{fggbar} of Appendix~\ref{app:solpert} for
the perturbative expansion of the metric $f_{\mu\nu}$. Keeping only
the leading non-linear correction we obtain (recall that we choose
$\alpha=1$ for this application)
\be{fsol2} f_{\mu\nu} = \frac{1}{2}\left(g_{\mu\nu} +
\ul{g}_{\mu\nu}\right) - \frac{1}{2}f^{\rho\sigma}\,h_{\mu\rho}
h_{\nu\sigma} + \calO\left(h^4\right)\,,\ee
where we remind that $h_{\mu\nu} = \frac{1}{2}(g_{\mu\nu} -
\ul{g}_{\mu\nu})$ by definition. The non-linear correction plays a
crucial role for the 1PN limit as it rules the value of the PPN
parameter $\beta$~\cite{Will}. Actually it happens that the elegant
prescription~\eqref{fdef} we have adopted for the metric $f_{\mu\nu}$
yields the correct value for the parameter $\beta$. Working out
Eq.~\eqref{fsol2} at 1PN order we find that the 1PN parametrization of
the metric $f_{\mu\nu}$ is simply obtained from the half sum of the
1PN potentials parametrizing the two metrics $g_{\mu\nu}$ and
$\ul{g}_{\mu\nu}$, namely
\be{f1PN} f^\text{1PN}_{\mu\nu} = f_{\mu\nu}\Bigl[\frac{V+\ul{V}}{2},
  \frac{V_i+\ul{V}_i}{2}\Bigr]\,.\ee
The 1PN metrics being properly parametrized, we insert them into the
action~\eqref{actionSS} and vary it with respect to $V$, $V_i$,
$\ul{V}$ and $\ul{V}_i$. We thus obtain two equations for $V$ and
$\ul{V}$ valid at order 1PN, which can be re-arranged into
[extending~\eqref{eqUUbar} to 1PN order]
\bes{eqVV} &&\Delta V +\frac{1}{c^2}\Bigl(3\partial^2_{t} V +
4\partial_{t}\partial_{i} V_{i}\Bigr) = - \frac{4\pi}{1+\varepsilon}
\Bigl[\bigl(1+2\varepsilon\bigr) \sigma_\text{b} +
  2\varepsilon\sigma_0 \Bigr]\,,\\ && \Delta\bigl(V+\ul{V}\bigr) +
\frac{1}{c^2}\Bigl(3\partial^2_{t} \bigl(V+\ul{V}\bigr) +
4\partial_{t}\partial_{i} \bigl(V_{i}+\ul{V}_{i}\bigr)\Bigr) = -
\frac{8\pi\varepsilon}{1+\varepsilon} \bigl(\sigma_\text{b} +
2\sigma_0\bigr) \,.\label{eqVVb}\ees
Similarly we obtain two equations for $V_i$ and $\ul{V}_i$,
\bes{eqVVi} &&\Delta V_i -
\partial_{i}\bigl(\partial_{t}V+\partial_{j}V_{j}\bigr) = -
\frac{4\pi}{1+\varepsilon} \Bigl[\bigl(1+2\varepsilon\bigr)
  \sigma^i_\text{b} + 2\varepsilon\,\sigma^i_0 \Bigr]\,,\\ &&
\Delta\bigl(V_i+\ul{V}_i\bigr) -
\partial_{i}\Bigl(\partial_{t}\bigl(V+\ul{V}\bigr)+\partial_{j}\bigl(V_j+\ul{V}_j\bigr)\Bigr)
= - \frac{8\pi\varepsilon}{1+\varepsilon} \bigl(\sigma^i_\text{b} +
2\sigma^i_0\bigr) \,,\ees
valid only at Newtonian order. The matter sources in these equations
are defined from the stress-energy tensor of the baryons as
\be{sigmab} \sigma_\text{b} =
\frac{T_\text{b}^{00}+T_\text{b}^{ii}}{c^2}\,,\qquad \sigma^i_\text{b}
= \frac{T_\text{b}^{0i}}{c}\,.\ee
These definitions are also valid if one includes some internal energy
and pressure into the baryonic part of the action~\eqref{actionSS}. A
1PN order we obtain for the matter sources
\be{sigmabexpl} \sigma_\text{b} = \rho^*_{b}\biggl(1-\frac{V}{c^2} +
\frac{3}{2}\frac{\bm{v}_{b}^2}{c^2}\biggr) \,,\qquad \sigma^i_\text{b}
= \rho^*_{b}\,v_{b}^{i} \,,\ee
which can easily be generalized to the case when adding internal
energy and pressure. Similarly we have posed for the dark matter,
\be{sigma0} \sigma_0 = \frac{T_0^{00}+T_0^{ii}}{c^2}\,,\qquad
  \sigma^i_0 = \frac{T_0^{0i}}{c}\,.\ee

Like in Sec.~\ref{sec:NRlimit} the relevant physics of our model is
the limiting case where $\varepsilon\ll 1$. Applying this limit on
Eqs.~\eqref{eqVV}--\eqref{eqVVi} we obtain the equations for the 1PN
potentials parametrizing the ordinary metric $g_{\mu\nu}$ felt by the
baryons as
\be{eqVViGR} \Box V = - 4\pi\,\sigma_\text{b}\,,\qquad \Delta V_i = -
4\pi\,\sigma^i_\text{b}\,,\ee
where we have used the harmonic coordinate condition in the ordinary
sector ${\partial_{t} V + \partial_{i} V^i = \calO\bigl( c^{-2}
  \bigr)}$, with the potentials in the dark sector being given by
$\ul{V}=-V$ and $\ul{V}_i=-V_i$. As the equations~\eqref{eqVViGR} are
the same as the standard equations of the 1PN limit of GR, see
\textit{e.g.}  Ref.~\cite{Bliving14}, we conclude that the model has
the same 1PN limit as GR and is therefore viable in the SS. One can
check directly from Eqs.~\eqref{eqVViGR} that all the PPN parameters
of the theory agree with their GR values~\cite{Will}. We emphasize
again that the PN limit works thanks to our particular
prescription~\eqref{fdef} for defining the interaction metric
$f_{\mu\nu}$ in the original action. Indeed the non-linear term coming
from that prescription [see Eq.~\eqref{fsol2}], turns out to be
exactly the one necessary to ensure that $\beta^\text{PPN}=1$.

To fully support the latter conclusions, let us look in more detail
at the fate of the residual dark matter contributions in
Eqs.~\eqref{eqVV}--\eqref{eqVVi}. Indeed, when taking the limit
$\varepsilon\to 0$ one must be careful with the fact that the
effective dark matter observed in cosmology has been found to be
$\varepsilon$ times the ``bare'' dark matter, see Eqs.~\eqref{rhoM}
or~\eqref{effdensity} with $\alpha=1$. Posing thus $\sigma_\text{DM} =
2\varepsilon\sigma_0$ and $\sigma^i_\text{DM} =
2\varepsilon\sigma^i_0$ we could expect that there should be some
remaining dark matter terms $\sigma_\text{DM}$ and
$\sigma^i_\text{DM}$ in the right-hand sides
of~\eqref{eqVViGR}. Similarly, we could expect the presence of a
  residual dark matter contribution
  $\rho_\text{DM}^*=2\varepsilon\rho^*$ in the right-side of the MOND
  equation, see~\eqref{Upol} or~\eqref{BMeq}.

However we now argue that this dark matter is negligible with respect
to baryonic matter, so that we can blindly apply the limit
$\varepsilon\to 0$ as we did to obtain~\eqref{eqVViGR}. This is due to
a property of ``\textit{weak clustering of dipolar dark matter}''
which is at work in the present model. According to this property the
dark matter medium should not cluster much during the cosmological
evolution, so that the dark matter density contrast in a typical
galaxy at low redshift after a long cosmological evolution should be
smaller than the density contrast of baryonic matter. In the present
model this property is the consequence of the fact that the dipolar
dark matter particles obey the geodesic equation $a_0^\mu=0$ with
respect to the metric $f_{\mu\nu}$,\,\footnote{Indeed, the
  acceleration $a_0^\mu$ is given by Eq.~\eqref{a0mu} where we recall
  that the parameter $\alpha$ is very close to one in the physically
  relevant case $\varepsilon\to 0$, \textit{i.e.}
  $\alpha=1+\calO(\varepsilon)$ from Eq.~\eqref{limitalpha}.} while
the baryons obey the geodesic equation $a_\text{b}^\mu=0$ with respect
to the ordinary metric $g_{\mu\nu}$. Therefore the baryons are
accelerated relatively to the dark matter medium. Using the result
that in the limit $\varepsilon\to 0$ the metric $f_{\mu\nu}$ is almost
flat, we see that $a_0^\mu=0$ implies that the dark matter fluid is
unaccelerated in the ordinary three-dimensional sense with respect to
some averaged cosmological matter distribution. In the Newtonian
approximation we have indeed seen that $\ud \bm{v}_0/\ud t =
\frac{1}{2}\bm{\nabla}(U+\ul{U}) = 0$. We thus expect that
$\sigma_\text{DM}$ and $\sigma^i_\text{DM}$ (or
  $\rho_\text{DM}^*$ in the MOND equation) will be negligible
compared to the baryonic contributions in generic galaxies and
in the solar system, and may even take very small typical average
cosmological values, \textit{e.g.}  $\sigma_\text{DM}\sim
10^{-29}\,\mathrm{g}\,\mathrm{cm}^{-3}$. The property of weak
clustering of dark matter in the present model\,\footnote{Recall that
  in the previous model of dipolar dark matter~\cite{BL08, BL09}, the
  ``weak clustering of dipolar dark matter'' was used as an
  \textit{hypothesis} but not as a property logically deduced within
  that model.} could be checked by implementing numerical $N$-body
cosmological simulations.

\section{Conclusion}
\label{sec:concl}

In this paper we have shown how a specific form of dark matter, made
of two different species of particles coupled to two different
metrics, and interacting through a specific internal force field,
could permit to interpret in the most natural way the phenomenology of
MOND by a mechanism of gravitational polarization. In this approach
the dark matter medium appears as a polarizable plasma-like fluid of
space-like dipole moments, aligned with the local gravitational field
generated by ordinary baryonic matter. On the other hand, that
particular form of dark matter reproduces the cosmological model
$\Lambda$-CDM at first order cosmological perturbations, and is thus
consistent with the observed spectrum of anisotropies of the
CMB~\cite{HuD02}. Furthermore we have shown that the theory is viable
in the solar system as it predicts the same PPN parameters as
GR. Finally the gravitational sector of the model is consistent
(ghost-free) at linear order around a Minkowski background.

Improvements with respect to the previous model of dipolar dark
matter~\cite{BL08, BL09} include the hypothesis of ``weak clustering
of dipolar dark matter'' which is probably built in the model, and the
fact that the dark matter medium is stable, as it undergoes stable
plasma-like oscillations when analyzed in perturbations. Another
important feature of the present model is that the mechanism of
alignment of the polarization with the gravitational field, and
consequently the validity of the MOND equation \textit{stricto sensu},
is expected to hold in any non static and non spherical cases. This is
important because it has been shown that MOND works well in describing
the highly dynamical evolution and collision of galaxies~\cite{TC07,
  TC08, GFC07} and the non-spherical polar ring structures of
galaxies~\cite{LFK13}.

On the other hand, while Refs.~\cite{BL08, BL09} investigate a
pure model of modified dark matter in standard GR, the present model
is less economical in that it postulates both a non standard form of
dark matter and a modification of gravity in the form of a bimetric
extension of GR. Such compromise between dark matter and modified
gravity is perhaps the price to pay for reconciling within a single
relativistic framework the conflicting observations of dark matter at
large cosmological scales and at small galactic scales. It would be
very interesting to test the model by performing $N$-body cosmological
numerical simulations, and notably to investigate the intermediate
scale of galaxy clusters at which the pure modified gravity theories
generally meet problems~\cite{FamMcG12}.


\section*{Acknowledgements}
It is a pleasure to thank Gilles Esposito-Far\`ese, Benoit Famaey,
Alexandre Le Tiec and Moti Milgrom for interesting discussions and
especially for very useful remarks on a preliminary version of this
work. We acknowledge partial support from Agence Nationale de la
Recherche \textit{via} the Grant THALES (ANR-10-BLAN-0507-01-02).

%
%
\appendix

\section{Perturbative solution for the metric $ f_{\mu\nu} $}
\label{app:solpert}

In this Appendix we find the perturbative solution of the implicit
definition~\eqref{fdef} of the metric $f_{\mu\nu}$ given the two
metrics $g_{\mu\nu}$ and $\ul{g}_{\mu\nu}$, namely
\be{fdefapp} f^{\rho\sigma} g_{\rho\mu}\,\ul{g}_{\nu\sigma} =
f^{\rho\sigma} g_{\rho\nu}\,\ul{g}_{\mu\sigma} = f_{\mu\nu} \,.\ee
Let us first gain an insight into the meaning of this prescription by
looking at the solution in terms of matrices. For this purpose we pose
$G^\nu_\mu = f^{\nu\rho}g_{\mu\rho}$ and $\ul{G}^\nu_\mu =
f^{\nu\rho}\ul{g}_{\mu\rho}$, and define the associated two matrices
$G = (G^\nu_\mu)$ and $\ul{G} = (\ul{G}^\nu_\mu)$. With such a matrix
notation the relation~\eqref{fdefapp} becomes, with
$\id=(\delta^\nu_\mu)$ denoting the unit matrix:
\be{matrix} G \ul{G} = \ul{G} G = \id\,,\ee
which means that $\ul{G}$ is the inverse of $G$. 

Next we look for the solution of Eqs.~\eqref{matrix} in the form of
the perturbative expansion
\be{GGbarmat} G = \alpha\bigl(\id + H + X\bigr)\,,\qquad \ul{G} =
\frac{1}{\alpha}\bigl(\id - H + X\bigr) \,,\ee
where $\alpha$ denotes a constant, the matrix $H$ represents the first
order perturbation and is defined by $H = \frac{1}{2}(\alpha^{-1}G -
\alpha\,\ul{G})$, and the matrix $X$ admits an expansion series in
powers of $H$ starting at the \textit{second} order in $H$. The matrix
equation to be solved is found to be $X^2+2X-H^2=0$, whose appropriate
solution reads $X=-\id+\sqrt{\id + H^2}$, where we have defined the
matrix $\sqrt{\id + H^2}$ by its expansion series in powers of $H$,
that is
\be{sqrtmat} \sqrt{\id + H^2} = \sum_{p=0}^{+\infty}
\gamma_p\,H^{2p}\quad\text{with}\quad\gamma_p =
\frac{(-)^{p+1}(2p-3)!!}{2^p p!}\,.\ee
It is interesting to note that the same expansion series plays a
crucial role in the definition of the mass term in resummed ghost-free
massive gravity theories, see \textit{e.g.}~\cite{deRham11}. Finally
our perturbative solution is
\bes{gsolmat} G &=& \alpha\Bigl(H + \sqrt{\id + H^2}\Bigr)\,,\\ \ul{G}
&=& \frac{1}{\alpha}\Bigl(- H + \sqrt{\id + H^2}\Bigr)\,.\ees
Notice that such a perturbative solution $G$ obviously commutes with
$\ul{G}$ and therefore only one out of the two
equations~\eqref{matrix} is sufficient.

Having the above solution in hands we conveniently lower back the
contravariant index so as to restore the metrics in a standard
form. The expansion variable is ${h_{\mu\nu} = H^\rho_\mu f_{\rho\nu}
  = \frac{1}{2}(\alpha^{-1}g_{\mu\nu} - \alpha\ul{g}_{\mu\nu})}$ which
was used as the metric perturbation in Sec.~\ref{sec:displ}. The
solution reads
\be{gsol} g_{\mu\nu} = \alpha\Bigl(f_{\mu\nu} + h_{\mu\nu} +
x_{\mu\nu}\Bigr) \,,\qquad \ul{g}_{\mu\nu} =
\frac{1}{\alpha}\Bigl(f_{\mu\nu} - h_{\mu\nu} + x_{\mu\nu}\Bigr)
\,,\ee
where $x_{\mu\nu} = X^\rho_\mu f_{\rho\nu}$ is at least of second
order and is given by
\be{Xsol} x_{\mu\nu} = \sum_{p=1}^{+\infty}
\gamma_p\,H_{\mu}^{\rho_1}H_{\rho_1}^{\rho_2}\cdots
H_{\rho_{2p-2}}^{\rho_{2p-1}}h_{\nu\rho_{2p-1}}\,.\ee
In particular $f_{\mu\nu}$ can be determined from the two metrics
$g_{\mu\nu}$ and $\ul{g}_{\mu\nu}$ by the relation
\be{fggbar} f_{\mu\nu} = \frac{1}{2}\left(\alpha^{-1}g_{\mu\nu} +
\alpha\,\ul{g}_{\mu\nu}\right) - \sum_{p=1}^{+\infty}
\gamma_p\,H_{\mu}^{\rho_1}H_{\rho_1}^{\rho_2}\cdots
H_{\rho_{2p-2}}^{\rho_{2p-1}}h_{\nu\rho_{2p-1}}\,,\ee
which is nevertheless implicit because
$H^\rho_\mu=f^{\rho\sigma}h_{\mu\sigma}=\frac{1}{2}f^{\rho\sigma}(\alpha^{-1}
g_{\mu\sigma} - \alpha \ul{g}_{\mu\sigma})$ still depends on
$f^{\rho\sigma}$. The first non-linear correction term in
Eq.~\eqref{fggbar} plays an important role when investigating the 1PN
limit of the theory in Sec.~\ref{sec:1PNlimit}.

Finally we can vary Eq.~\eqref{fggbar} with respect to $g_{\mu\nu}$
and $\ul{g}_{\mu\nu}$ to determine perturbatively (\textit{i.e.} order
by order) the tensorial coefficients
$\mathcal{A}_{\,\mu\nu}^{\,\rho\sigma}$ and
$\ul{\mathcal{A}}_{\,\mu\nu}^{\,\rho\sigma}$ defined in
Eq.~\eqref{deltaf} as
\be{deltafapp} \delta f_{\mu\nu} =
\frac{1}{2}\,\mathcal{A}_{\,\mu\nu}^{\,\rho\sigma}\,\delta
g_{\rho\sigma} +
\frac{1}{2}\,\ul{\mathcal{A}}_{\,\mu\nu}^{\,\rho\sigma}\,\delta
\ul{g}_{\rho\sigma}\,.\ee
From Eqs.~\eqref{fdefapp} we find that such coefficients must obey the
equations
\bes{relAAbar} \frac{1}{2}\bigl(\mathcal{A}_{\,\mu\nu}^{\,\rho\sigma}
+ G_{\mu}^\lambda
\,\ul{G}_{\nu}^\tau\,\mathcal{A}_{\,\lambda\tau}^{\,\rho\sigma}\bigr)
&=& \delta_{\mu}^{(\sigma}\,\ul{G}_{\nu}^{\rho)}
\,,\\ \frac{1}{2}\bigl(\ul{\mathcal{A}}_{\,\mu\nu}^{\,\rho\sigma} +
G_{\mu}^\lambda
\,\ul{G}_{\nu}^\tau\,\ul{\mathcal{A}}_{\,\lambda\tau}^{\,\rho\sigma}\bigr)
&=& G_{\mu}^{(\rho}\,\delta_{\nu}^{\sigma)} \,,\ees
together with the same equations with $\mu$ and $\nu$ exchanged. These
equations can be solved iteratively to any order. For instance we find
the solutions up to second order as
\bes{solAAbar2} \mathcal{A}_{\,\mu\nu}^{\,\rho\sigma} &=&
\frac{1}{\alpha}\Bigl(\delta_{(\mu}^\rho \,\delta_{\nu)}^\sigma -
H_{(\mu}^{(\rho} \,\delta_{\nu)}^{\sigma)} + \frac{1}{2}H_{(\mu}^\rho
\,H_{\nu)}^\sigma \Bigr) +
\calO\left(3\right)\,,\\ \ul{\mathcal{A}}_{\,\mu\nu}^{\,\rho\sigma}
&=& \alpha\Bigl(\delta_{(\mu}^\rho \,\delta_{\nu)}^\sigma +
H_{(\mu}^{(\rho} \,\delta_{\nu)}^{\sigma)} + \frac{1}{2}H_{(\mu}^\rho
\,H_{\nu)}^\sigma \Bigr) + \calO\left(3\right)\,.\ees One can check
that the relations $f^{\mu\nu}\mathcal{A}_{\,\mu\nu}^{\,\rho\sigma} =
g^{\rho\sigma}$ and
$f^{\mu\nu}\ul{\mathcal{A}}_{\,\mu\nu}^{\,\rho\sigma} =
\ul{g}^{\rho\sigma}$, which are direct consequences of
$f^2=g\,\ul{g}$, are satisfied to this order.

\section{Plasma-like hypothesis}
\label{app:linearorder}

The dark matter fluids are described by the conserved currents
  $j^\mu$ and $\ul{j}^\mu$ defined by Eqs.~\eqref{jmu}. Here we
  implement the idea that they perturbatively differ from a single
  equilibrium fluid described by the current $j_0^\mu=\rho_0 u_0^\mu$,
  such that $f_{\mu\nu}u_0^\mu u_0^\nu=-1$ and $\calD_\mu j_0^\mu=0$.
To do that, suppose for simplicity that the equilibrium fluid is made
of particles with coordinate density ${\rho_0^*(\mathbf{x},t)=\sum_A
  m_A\delta[\mathbf{x}-\bm{x}_A(t)]}$ (with $\delta$ being the usual
three-dimensional Dirac function), satisfying the usual continuity
equation $\partial_t \rho_0^* +
\partial_i(\rho_0^* v_0^i) = 0$, where $v_0^i(\mathbf{x},t)$ is the
Eulerian velocity field. Then the coordinate density of the displaced
fluid is defined with respect to that of the equilibrium fluid as
$\rho^*(\mathbf{x},t)=\sum_A
m_A\delta[\mathbf{x}-\bm{x}_A(t)-\bm{y}_A(t)]$, where $y_A^i(t)$ is
the displacement of the particles' positions. Introducing the Eulerian
displacement field $y^i(\mathbf{x},t)$ associated with
$y_A^i(t)$, we find that ${\rho^*=\rho_0^*-\partial_i(\rho_0^* y^i)}$ to
first order in the displacement, while the coordinate velocity reads
${v^i=v^i_0+\frac{\ud y^i}{\ud t} - y^j\partial_j v_0^i}$, where
$\ud/\ud t$ is the convective derivative. Introducing the coordinate
current $J_*^0=\rho^*$ and $J_*^i=\rho^* v^i$ such that $\partial_\mu
J_*^\mu = 0$, and two displacement vectors $y^\mu$ and $\ul{y}^\mu$
for the two fluids, we obtain\,\footnote{Note that one can always
  choose $y^0=\ul{y}^0=0$ to define the two displacement four-vectors
  $y^\mu$ and $\ul{y}^\mu$.}
\bes{jjbar*} J_*^\mu &=& J_{0*}^\mu + \partial_\nu\left(J_{0*}^\nu
y^\mu-J_{0*}^\mu y^\nu\right) + \calO\left(2\right)\,,\\ \ul{J}_*^\mu
&=& J_{0*}^\mu + \partial_\nu\left(J_{0*}^\nu \ul{y}^\mu-J_{0*}^\mu
\ul{y}^\nu\right) + \calO\left(2\right)\,.\ees
In what follows we systematically work at first order in the
displacement vectors $y^\mu$ and $\ul{y}^\mu$, and assume that their
gradients are numerically of the same order as the metric perturbation
$h_{\mu\nu}$, namely that $\nabla y\sim \ul{\nabla}\ul{y}\sim h =
\calO(1)$, so that the remainders $\calO(2)$ in Eqs.~\eqref{jjbar*}
are of the same order as those in Eqs.~\eqref{ggbar}. The
expressions~\eqref{jjbar*} are covariantized in the usual way by
defining $j^\mu=J_*^\mu/\sqrt{-f}$, \textit{etc.}, and we obtain (see
\textit{e.g.}~\cite{Taub54})
\bes{jjbarann} j^\mu &=& j_0^\mu + \calD_\nu\left(j_0^\nu
y_\perp^\mu-j_0^\mu y_\perp^\nu\right) +
\calO\left(2\right)\,,\\ \ul{j}^\mu &=& j_0^\mu +
\calD_\nu\left(j_0^\nu \ul{y}_\perp^\mu-j_0^\mu
\ul{y}_\perp^\nu\right) + \calO\left(2\right)\,.\ees
We have taken advantage of the structure of the terms to replace the
displacement vectors by their projections perpendicular to the
four-velocity of the equilibrium fluid, namely
${y_\perp^\mu=\perp^\mu_\nu y^\nu}$ and $\ul{y}_\perp^\mu=\perp^\mu_\nu
\ul{y}^\nu$ where $\perp^{\mu\nu}\equiv f^{\mu\nu}+u_0^\mu u_0^\nu$.
Coming back to the scalar densities $\rho=\sqrt{-g_{\mu\nu}J^\mu
  J^\nu}$ and $\ul{\rho}=\sqrt{-\ul{g}_{\mu\nu}\ul{J}^\mu\ul{J}^\nu}$,
taking into account the relations~\eqref{jmu} between currents and
using at first order $\beta=\alpha^2[1+\frac{h}{2}+\calO(2)]$ and
$\ul{\beta}=\alpha^{-2}[1-\frac{h}{2}+\calO(2)]$, where $h \equiv
f^{\mu\nu}h_{\mu\nu}$, we obtain
\bes{rhorho0} \rho &=& \alpha^{-3/2}\biggl[\rho_0 \Bigl(1 -
  \frac{h}{2} - \frac{1}{2} h_{\mu\nu} u_0^\mu u_0^\nu + a_{0\mu}
  y_\perp^\mu \Bigr) - \calD_\mu
  \bigl(\rho_0\,y_\perp^\mu\bigr)\biggr] + \calO\left(2\right)
\,,\\ \ul{\rho}&=& \alpha^{3/2}\biggl[\rho_0 \Bigl(1 + \frac{h}{2} +
  \frac{1}{2} h_{\mu\nu} u_0^\mu u_0^\nu + a_{0\mu} \ul{y}_\perp^\mu
  \Bigr) - \calD_\mu \bigl(\rho_0\,\ul{y}_\perp^\mu\bigr)\biggr] +
\calO\left(2\right)\,,\ees
where $a_0^\mu \equiv u_0^\nu\calD_\nu u_0^\mu$ is the acceleration of
the equilibrium configuration, and $a_{0\mu}=f_{\mu\nu}a_0^\nu$. For
the four-velocities we get
\bes{umuu0} u^\mu &=& \alpha^{-1/2}\biggl[u_0^\mu \Bigl(1 +
  \frac{1}{2} h_{\rho\sigma} u_0^\rho u_0^\sigma - a_{0\mu}
  y_\perp^\mu \Bigr) + \mathscr{L}_{u_0}y_\perp^\mu\biggr] +
\calO\left(2\right)\,,\\ \ul{u}^\mu &=& \alpha^{1/2}\biggl[u_0^\mu
  \Bigl(1 - \frac{1}{2} h_{\rho\sigma} u_0^\rho u_0^\sigma - a_{0\mu}
  \ul{y}_\perp^\mu \Bigr) + \mathscr{L}_{u_0}\ul{y}_\perp^\mu\biggr] +
\calO\left(2\right)\,,\ees
in which we made use of the Lie derivative, \textit{e.g.}
$\mathscr{L}_{u_0}y_\perp^\mu = u_0^\nu \calD_\nu y_\perp^\mu -
y_\perp^\nu \calD_\nu u_0^\mu$.

\section{Linearisation around a Minkoswki background}
\label{app:lineargravact}

In this Appendix we derive the gravitational part $S_{g}$ of the
  action~\eqref{action} at quadratic order in perturbation around a
  Minkowski background. Ignoring for simplicity the cosmological
  constants, we thus start from
\be{Sg} S_{g} = \frac{1}{32\pi} \int\ud^{4}x \left\{ \sqrt{-g}\,R +
\sqrt{-\ul{g}}\,\ul{R} +
\frac{2}{\varepsilon}\sqrt{-f}\,\mathcal{R}\right\} \,, \ee 
where the interaction metric $f_{\mu\nu}$ is defined from the two
metrics $g_{\mu\nu}$, $\ul{g}_{\mu\nu}$ by the
prescription~\eqref{fdef}. To linear order we have
$g_{\mu\nu}=\eta_{\mu\nu}+k_{\mu\nu}+\calO(2)$,
$\ul{g}_{\mu\nu}=\eta_{\mu\nu}+\ul{k}_{\mu\nu}+\calO(2)$ and
$f_{\mu\nu}=\eta_{\mu\nu}+s_{\mu\nu}+\calO(2)$, where $\eta_{\mu\nu}$
is the Minkowski metric and
\be{smunu} s_{\mu\nu} =
\frac{1}{2}\bigl(k_{\mu\nu}+\ul{k}_{\mu\nu}\bigr) \,.  \ee 
With these notations the variable $h_{\mu\nu}$ defined by
Eq.~\eqref{ggbar} (with $\alpha=1$) reads
\be{hmunu} h_{\mu\nu} =
\frac{1}{2}\bigl(k_{\mu\nu}-\ul{k}_{\mu\nu}\bigr) \,. \ee 
It is now straightforward to derive the quadratic part of the action
in terms of the two variables~\eqref{smunu}-~\eqref{hmunu}. We find
that the two sectors associated with those variables decouple from
each other, namely 
\be{quadract} S_{g} = \frac{1}{32\pi} \int\ud^{4}x \left\{
-\frac{1}{2}\,\partial_{\mu}h_{\nu\rho}\,\partial^{\mu}\hat{h}^{\nu\rho}
+\hat{H}_{\mu}\hat{H}^{\mu} +
\frac{1+\varepsilon}{\varepsilon}\Bigl(-\frac{1}{2}\,
\partial_{\mu}s_{\nu\rho}\,\partial^{\mu}\hat{s}^{\nu\rho}
+\hat{S}_{\mu}\hat{S}^{\mu}\Bigr) \right\} +\calO(3) \,, \ee
where we define
$\hat{h}^{\mu\nu}=h^{\mu\nu}-\tfrac{1}{2}\eta^{\mu\nu}h$,
$\hat{H}^{\mu}=\partial_{\nu}\hat{h}^{\mu\nu}$ and similarly for
$\hat{s}^{\mu\nu}$ and $\hat{S}^{\mu}$. Thus the action appears at
that order as the sum of two massless non-interacting spin-2 fields,
with positive sign in the case where $\varepsilon>0$. Since this
action enjoys two reparametrization invariances $\delta
h_{\mu\nu}=2\partial_{(\mu}\xi_{\nu)}$ and $\delta
s_{\mu\nu}=2\partial_{(\mu}\chi_{\nu)}$, where $\xi_{\nu}$ and
$\chi_{\nu}$ are two independent functions, each spin-2 field
propagates only two degrees of freedom as expected for massless
gravitons~\cite{Boul.al.00}. However the full action of the model
should still be investigated at the non-linear level for which the
number of propagating gravitational modes should be investigated. This
question is addressed in Ref.~\cite{BH15}.

\section{Cosmological perturbations}
\label{app:cosmopert}

\subsection{Gravitational perturbations}
\label{sec:gravpert}

We assume that both metrics $g_{\mu\nu}$ and $\ul{g}_{\mu\nu}$,
which as we have seen in Sec.~\ref{sec:displ} differ perturbatively
from each other, take the form of a linear perturbation around the
FLRW background~\eqref{FLRW}. The metric intervals read
then,\,\footnote{In this section we omit indicating that second order
  perturbations $\calO(2)$ are systematically neglected. Our notation
  $\tilde{h}_{i}$, $\tilde{\ul{h}}_{\,i}$ and $\tilde{h}_{ij}$,
  $\tilde{\ul{h}}_{\,ij}$ is to avoid confusion with the components of
  the covariant tensor $h_{\mu\nu}$.}
\be{ds2} \text{d}s^2 = a^2\left[-(1+2A)\,\text{d}\eta^2 +
  2\,\tilde{h}_{i}\,\text{d}\eta\,\text{d}x^{i} +
  (\gamma_{ij}+\tilde{h}_{ij})\,\text{d}x^{i}\,\text{d}x^{j} \right]
\,,\ee
and similarly for the other metric interval $\text{d}\ul{s}^{\,2}$,
\be{ds2bar} \text{d}\ul{s}^2 =
\ul{a}^2\left[-(1+2\ul{A})\,\text{d}\eta^2 +
  2\,\tilde{\ul{h}}_{i}\,\text{d}\eta\,\text{d}x^{i} +
  (\gamma_{ij}+\tilde{\ul{h}}_{ij})\,\text{d}x^{i}\,\text{d}x^{j}
  \right] \,.\ee
The variables $A$, $\tilde{h}_{i}$, $\tilde{h}_{ij}$ and $\ul{A}$,
$\tilde{\ul{h}}_{\,i}$, $\tilde{\ul{h}}_{\,ij}$ respectively denote
the metric perturbations for the metrics $g_{\mu\nu}$ and
$\ul{g}_{\mu\nu}$. An easy computation yields the perturbation of the
metric $f_{\mu\nu}$ as
\be{ds2f} \text{d}s^2_{\!f} =
a\ul{a}\left[-(1+A+\ul{A})\,\text{d}\eta^2 +
  (\tilde{h}_{i}+\tilde{\ul{h}}_{i})\text{d}\eta\,\text{d}x^{i} +
  (\gamma_{ij}+\tfrac{1}{2}\tilde{h}_{ij} +
  \tfrac{1}{2}\tilde{\ul{h}}_{ij})\,\text{d}x^{i}\,\text{d}x^{j}
  \right] \,.\ee
Next we perform the standard Scalar-Vector-Tensor (SVT) decomposition
of the metric perturbations (see~\cite{PUcosmo} for a review). For the
ordinary sector associated to $g_{\mu\nu}$ we pose
\bes{SVT} \tilde{h}_{i} &=& \uD_{i}\,B+B_{i} \,,
\\\tilde{h}_{ij} &=& 2\,C\,\gamma_{ij} + 2\uD_{i} \uD_{j}\,E +
2\,\uD_{(i}\,E_{j)} + 2\,E_{ij}\,,\ees
and identically for the dark sector $\ul{g}_{\mu\nu}$. All spatial
indices are raised and lowered with $\gamma_{ij}$ and its inverse
$\gamma^{ij}$. The vectors $B^{i}$, $E^{i}$, $\ul{B}^{\,i}$,
$\ul{E}^{\,i}$ defined in this way are divergenceless, while the
second-rank tensors $E^{ij}$, $\ul{E}^{\,ij}$ are divergenceless and
traceless,
\bes{SVTcond} && \uD_{i}\,B^{i} = \uD_{i}\,E^{i} =
\uD_{i}\,\ul{B}^{\,i} = \uD_{i}\,\ul{E}^{\,i} = 0 \,,\\ &&
\uD_{j}\,E^{ij} = E^{i}_{\phantom{i}i} = \uD_{j}\,\ul{E}^{\,ij} =
\ul{E}^{\,i}_{\phantom{\,i}i} =0 \,. \ees
As usual one can construct gauge-invariant quantities from these
variables~\cite{PUcosmo}. We shall use in the ordinary sector,
\bes{invgauge} \Phi &=& A+B'-E''+\mathcal{H}\bigl(B-E'\bigr)\,,\\ \Psi
&=& -C-\mathcal{H}\bigl(B-E'\bigr) \,,\\ X &=&
A-C-\bigl(C/\mathcal{H}\bigr)' \,,\\ \Phi_{i} &=& E'_{i}-B_{i}
\,. \ees
Note that the scalar $X$ so defined is not independent from the two
other scalars $\Phi$ and $\Psi$,
\be{Xdef} X = \Phi+\Psi+\biggl(\frac{\Psi}{\mathcal{H}}\biggr)'\,. \ee
Note also that $E_{ij}$ is already a gauge invariant quantity. The
same definitions apply of course to the dark sector
$\ul{g}_{\mu\nu}$, for the gauge-invariant quantities $\ul{\Phi}$,
$\ul{\Psi}$, $\ul{X}$, $\ul{\Phi}_i$ and $\ul{E}_{ij}$.
From Eqs.~\eqref{ggbar} we see that the second-rank tensor field
$h_{\mu\nu}=\frac{1}{2}(\alpha^{-1}g_{\mu\nu}-\alpha\ul{g}_{\mu\nu})$
can be written
as
\bes{hSVT} h_{00} &=& -a\ul{a} \,dA\,,\\ h_{0i} &=& \frac{a\ul{a}}{2}
\,d \tilde{h}_{i} = \frac{a\ul{a}}{2}\,\bigl(\uD_{i}\,dB+ dB_{i}\bigr)
\,,\\ h_{ij} &=& \frac{a\ul{a}}{2} \,d\tilde{h}_{ij} =
a\ul{a}\,\bigl(d C\,\gamma_{ij} + \uD_{i} \uD_{j}\,dE +
\uD_{(i}\,dE_{j)} + d E_{ij}\bigr)\,, \ees
where for any spatial scalar, vector or tensor $P$ (gauge-invariant or
not) we denote the \textit{difference} between $P$ in the ordinary
sector and the corresponding quantity $\ul{P}$ in the dark sector by
\be{dP} dP \equiv P-\ul{P}\,. \ee
It is evident that the difference of gauge invariant quantities is
gauge invariant, but notice that the difference of any quantities
(scalar, vector or tensor) \textit{is} a gauge invariant
quantity. Thus in the following we extensively use the fact that $dA$,
$dB$, $dC$, $dE$, $dB_{i}$, $dE_{i}$ and $dE_{ij}$ are gauge
invariant. In addition we have also at our disposal the differences of
gauge-invariant variables $d\Phi$, $d\Psi$, $dX$ and $d\Phi_i$ defined
similarly to Eqs.~\eqref{invgauge}.

\subsection{Matter perturbations}
\label{sec:matpert}

We have in our model three fluids, two fluids of dark matter described
by scalar densities $\rho$ and $\ul{\rho}$ and four-velocities $u^\mu$
and $\ul{u}^\mu$, and the fluid of baryons described by
$\rho_\text{b}$ and $u_\text{b}^\mu$. On the other hand, we have
learned from Sec.~\ref{sec:displ} [see
Eqs.~\eqref{jjbar} and also \eqref{rhorho0}--\eqref{umuu0}] how to relate the densities
and four-velocities of the two dark matter fluids \textit{via} an
auxiliary fluid described by $\rho_0$, $u_0^\mu$ corresponding to some
equilibrium configuration, and a space-like vector $\xi_\perp^\mu$
called the dipole moment.
In addition, we have the fluid associated with the cosmological
constant $\Lambda$. We already pointed out that in order to have a
true cosmological constant, even at first order in
perturbations (in agreement with the $\Lambda$-CDM model), we must
relate the three initial constants $\lambda$, $\ul{\lambda}$ and
$\lambda_f$ in the action~\eqref{action} in the way specified by
Eq.~\eqref{constcosmo}, and that $\Lambda$ then denotes the observed
cosmological constant.
At perturbative level the four-velocities of the two dark matter
fluids read ${u^\mu = \,\fond{u}^\mu + \delta u^\mu}$ and ${\ul{u}^\mu
  = \,\fond{\ul{u}}^\mu + \delta \ul{u}^\mu}$, with a similar notation
for the baryons. The background quantities are given in
Eqs.~\eqref{fonduubar}.
Recalling that the fluids $\rho$, $u^\mu$ and $\ul{\rho}$,
$\ul{u}^\mu$ are defined with respect to the metrics $g_{\mu\nu}$ and
$\ul{g}_{\mu\nu}$ respectively, their first-order perturbed velocities
read
\be{upert} u^\mu = \frac{1}{a}\bigl(1-A, \beta^i\bigr)\,,\qquad
\ul{u}^\mu = \frac{1}{\ul{a}}\bigl(1-\ul{A}, \ul{\beta}^i\bigr)\,. \ee
We perform the usual SVT decomposition,
\be{beta} \beta^i = \uD^i v + v^i\,,\qquad \uD_i v^i=0\,, \ee
and introduce the gauge invariant variables,
\bes{VVi} V &=& v + E'\,,\\ V^i &=& v^i + B^i\,. \ees
Obviously we have similar definitions for the dark sector,
\textit{e.g.} $\ul{V} = \ul{v} + \ul{E}'$, and for the baryons,
\textit{e.g.} $V_\text{b} = v_\text{b} + E'$. One can then express the
four-acceleration $a^\mu=u^\nu\nabla_\nu u^\mu$ in term of these
gauge-invariant quantities:
\be{amuamubar} a^\mu = \frac{1}{a^2}\left(0,
\uD^i\left(V'+\mathcal{H}V+\Phi\right)+ {V'}^i+\mathcal{H}
V^i\right)\,,\ee
and similarly for $\ul{a}^\mu=\ul{u}^\nu\ul{\nabla}_\nu
\ul{u}^\mu$. 
The scalar densities of dark matters read
$\rho=\,\fond{\rho}(1+\delta)$ and
$\ul{\rho}=\,\fond{\ul{\rho}}(1+\ul{\delta})$, where $\delta$ and
$\ul{\delta}$ are the density contrasts. We choose to express the
density contrasts in the ``flat slicing'' gauge (indicated by the
superscript F), defined by
\be{deltaF} \delta^\text{F} = \delta + 3 C \,,\qquad
\ul{\delta}^\text{F} = \ul{\delta} + 3 \ul{C} \,, \ee
and which obey the equations ($\Delta$ being the Laplacian associated
with the metric $\gamma_{ij}$)
\be{conteq} {\delta^\text{F}}' + \Delta V = 0\,,\qquad
   {\ul{\delta}^\text{F}}' + \Delta \ul{V} = 0\,. \ee
Similarly for the baryons, we define $\delta^\text{F}_\text{b} =
\delta_\text{b} + 3 C$ and get ${\delta^\text{F}}_{\!\!\text{b}}' +
\Delta V_\text{b} = 0$.
We now turn to the equilibrium configuration $\rho_0$, $u_0^\mu$
defined with respect to the metric $f_{\mu\nu}$, see
Eqs.~\eqref{jjbar} or~\eqref{rhorho0}--\eqref{umuu0}. The background
quantities have been given in~\eqref{backgrequil}. At linear order we
have $u_0^\mu = \,\fond{u}_0^\mu+\delta u_0^\mu$, which reads
explicitly
\be{u0} u_0^\mu =
\frac{1}{(a\ul{a})^{1/2}}\Bigl(1-\frac{1}{2}\bigl(A+\ul{A}\bigr),
\beta_0^i\Bigr)\,.\ee
The SVT decomposition and gauge-invariant variables proceed in the
same way,
\bes{SVT0} \beta_0^i &=& \uD^i v_0 + v_0^i\,,\qquad \uD_i
v_0^i=0\,,\\V_0 &=& v_0 + \frac{1}{2}\bigl(E'+\ul{E}'\bigr)\,,
\\ V_0^i &=& v_0^i + \frac{1}{2}\bigl(B^i+\ul{B}^i\bigr)\,.  \ees
For the scalar density we have $\rho_0=\,\fond{\rho}_0(1+\delta_0)$
and adopt the gauge invariant definition
\be{delta0F} \delta_0^\text{F} = \delta_0 +
\frac{3}{2}\bigl(C+\ul{C}\bigr)\,,\qquad {\delta_0^\text{F}}' + \Delta
V_0 = 0\,.\ee
Now the relations~\eqref{rhorho0}--\eqref{umuu0} translate immediately
to linear cosmological perturbations. With our choice of equilibrium
configuration the two displacement vectors read
$y^\mu=\frac{1}{2}\xi^\mu$ and $\ul{y}^\mu=-\frac{1}{2}\ul{\xi}^\mu$,
and the fluid at equilibrium obeys the equation of
motion~\eqref{a0mu}. Since $\xi^\mu_\perp =\perp^\mu_\nu\xi^\nu$ is
space-like it necessarily belongs to first order perturbations,
because a non-vanishing background dipole moment would break the
isotropy of space. Then the constraint
$\fond{u}_{0\mu}\xi_\perp^\mu=0$ implies $\xi_\perp^0=0$, so that we
have the SVT form,
\bes{zzi}\xi_\perp^\mu &=& \bigl(0,
\lambda^i\bigr)\,,\\ \text{with}\quad\lambda^i &=& \uD^i z +
z^i\,,\qquad \uD_i z^i =0\,,\ees
where $z$ and $z^i$ are by definition the SVT variables. Since the
background value is zero, they are directly gauge invariant. Using
Eqs.~\eqref{umuu0}, in which the acceleration $a_0^\mu$ can be
neglected since it is of first order [$a_0^\mu=\calO(1)$,
  see~\eqref{a0mu}], the variables $V$, $V^i$ and $\ul{V}$, $\ul{V}^i$
defined in~\eqref{VVi} are related to their partners $V_0$, $V_0^i$
by\,\footnote{In order to prove the following relations we used the
  useful formulae
\begin{eqnarray*} && h = dA+3dC+\Delta dE\,,\\ &&
h_{\mu\nu}\!\fond{u}_0^\mu\!\fond{u}_0^\nu = -dA\,.\end{eqnarray*} }
\bes{VVrel} V &=& V_0 + \frac{1}{2}\bigl(dE'+
z'\bigr)\,,\\ \ul{V} &=& V_0 - \frac{1}{2}\bigl(dE'+
z'\bigr)\,,\\ V^i &=& V_0^i + \frac{1}{2}\bigl(dB^i +
{z'}^i\bigr)\,,\\ \ul{V}^i &=& V_0^i - \frac{1}{2}\bigl(dB^i +
{z'}^i\bigr)\,.\ees
From Eqs.~\eqref{rhorho0} the corresponding gauge invariant density
contrasts are related by
\bes{deltarel} \delta^\text{F} &=& \delta_0^\text{F} -
\frac{1}{2}\Delta\bigl(dE + z\bigr)\,,\\\ul{\delta}^\text{F}
&=& \delta_0^\text{F} + \frac{1}{2}\Delta\bigl(dE + z\bigr)\,.\ees
Let us now deal with the dynamical equations of
motion~\eqref{DMeompert}, in which the four-accelerations in the SVT formalism
are given by \textit{e.g.}~\eqref{amuamubar}. Thus,
\bes{eomSVT} && V'+\mathcal{H}V+\Phi = -4\pi\fond{\rho} a^2
\,z\,,\\ && \ul{V}'+\mathcal{H}\ul{V} + \ul{\Phi} = 4\pi
\fond{\ul{\rho}} \ul{a}^2\,z\,, \\ && {V'}^i+\mathcal{H} V^i= -
4\pi\fond{\rho} a^2\,z^i \,,\\ && {\ul{V}'}^i+\mathcal{H} \ul{V}^i =
4\pi \fond{\ul{\rho}} \ul{a}^2\,z^i\,, \ees
with $\fond{\ul{\rho}} \ul{a}^2=\alpha\fond{\rho}
  a^2$. Similarly the equation of motion of the equilibrium fluid
found in Eq.~\eqref{a0mu} reads
\bes{eomSVT0} &&
V'_0+\mathcal{H}V_0+\frac{1}{2}\left(\Phi+\ul{\Phi}\right) =
-2\pi(1-\alpha)\!\fond{\rho} a^2 \,z\,, \\ && {V'}_0^i+\mathcal{H}
V_0^i = -2\pi(1-\alpha)\!\fond{\rho} a^2 \,z^i\,. \ees
Note that the latter equations are in fact implied by~\eqref{eomSVT}
when making use of the relations~\eqref{VVrel}. Finally, by computing
the differences $dV$ and $dV^i$ from Eqs.~\eqref{eomSVT}, and
using~\eqref{VVrel} together with the definition of $d\Phi$, we get
\bes{zzieq} z'' + \mathcal{H} z' + 4\pi(1+\alpha)\!\fond{\rho} a^2 \,z
&=& - dA-dB'-\mathcal{H}dB\,, \\ {z''}^i + \mathcal{H} {z'}^i +
4\pi(1+\alpha)\!\fond{\rho} a^2 \,z^i &=& - d{B'}^i -\mathcal{H}d
B^i\,, \ees
which constitute the SVT form of the equation of
evolution~\eqref{eqdipole} of the dipole moment. An alternative form
of these equations is provided in Appendix~\ref{app:pertEFEother},
see~\eqref{evolzzi}.

\section{Cosmological perturbations in the dark sector}
\label{app:pertEFEother}

In Sec.~\ref{sec:pertEFE} we investigated the cosmological
perturbations of the ordinary sector with metric $g_{\mu\nu}$. In this
Appendix we deal with the perturbation equations for the dark sector
with metric $\ul{g}_{\mu\nu}$. Actually it is simpler to consider the
equations for the \textit{differences} between the perturbation
variables in the two sectors. We shall prove that these equations
permit to determine all the variables in the model, even those which
cannot be measured by traditional cosmological observations taking
place in the ordinary sector.

We write the perturbation equations for the difference of the two
metrics in a way similar to Eqs.~\eqref{gravperteq}. We limit
ourselves to the three equations with sources since the other ones are
trivial. We get
\bes{gravperteqdark} &&\Delta d\Psi - 3\mathcal{H}^2 dX = 4 \pi
\,a^2\fond{\rho}\Bigl[-(p+q)\delta^\text{F}_\text{b} +
  p\,\delta^\text{F} + q\,\ul{\delta}^\text{F}+r\bigl(\Delta dE+
  dA\bigr)\Bigr]\,,\\ && d\Psi' + \mathcal{H} \,d\Phi =
-4\pi\,a^2\fond{\rho} \Bigl[-(p+q)V_\text{b} +p\,V + q\,\ul{V}
  +r\bigl(-dE' + \frac{1}{2}dB\bigr)\Bigr]\,,\\ &&(\Delta+
2K)\,d\Phi^i = - 16\pi\,a^2\fond{\rho} \Bigl[-(p+q)V^i_\text{b}
  +p\,V^i + q\,\ul{V}^i - \frac{1}{2}r\,dB^i\Bigr]\,,\ees
with coefficients
\begin{equation}\label{coeffs}\begin{aligned} p =
\displaystyle\frac{4\alpha(\varepsilon+\alpha)}{1+\alpha^2 +
  2\alpha\varepsilon}\,, \qquad q = \displaystyle
-\frac{4\alpha(1+\alpha\varepsilon)}{1+\alpha^2+2\alpha\varepsilon}\,,
\qquad r = \displaystyle
\frac{2\alpha(2\alpha+\varepsilon+\alpha^2\varepsilon)}
     {(\alpha+\varepsilon)(1+\alpha^2 +
       2\alpha\varepsilon)}\,.\end{aligned}
\end{equation}
From the right-hand sides of the equations~\eqref{gravperteqdark}, one
may define some effective variables for the matter fields, in a way
similar to~\eqref{effDM}.

Then the equations of continuity and of motion associated with these
matter variables are consequences of the equations themselves
(\textit{via} the Bianchi identities). The coefficients~\eqref{coeffs}
manage to simplify to give
\bes{dABBi} && d A' + \frac{1}{2}\Delta dB =
0\,,\label{eqdA}\\ &&\frac{1}{2}\Bigl(d B' + \mathcal{H} d B\Bigr) +
\,dA = -8\pi(\alpha
+\varepsilon)\,a^2\fond{\rho}\,z\,,\label{dAdBz}\\ &&
\frac{1}{2}\Bigl(d {B'}^i + \mathcal{H} d B^i\Bigr) = -8\pi(\alpha
+\varepsilon)\,a^2\fond{\rho} \,z^i\,.\label{dBiz}\ees
Gladly, we see that these equations guarantee consistency between
Eqs.~\eqref{conteq}--\eqref{eomSVT} and
Eqs.~\eqref{contDM}--\eqref{EOMDM}. Combining the two last
equations~\eqref{dAdBz}--\eqref{dBiz} with the equations of
motion~\eqref{zzieq}, we obtain two further equations for the
dipole moment $z$ and $z^i$,
\bes{evolzzi} z'' + \mathcal{H} z' +
4\pi\left(1-3\alpha-4\varepsilon\right)\fond{\rho}\,a^2\,z &=&
dA\,,\\ {z''}^i + \mathcal{H} {z'}^i +
4\pi\left(1-3\alpha-4\varepsilon\right)\fond{\rho}\,a^2\,z^i &=&
0\,.\ees
From all these equations we can determine $d A$, $d B$, $d B^i$ and
the dipole components $z$, $z^i$. Next, using $d\Phi^i=d {E'}^i-d B^i$
one gets from the vector and tensor mode differences
\bes{dEiij} && d {E''}^i + 2 \mathcal{H} d {E'}^i = d {B'}^i +
2\mathcal{H} d B^i\,,\\ && d {E''}^{ij} + 2 \mathcal{H} d {E'}^{ij} +
\left(2K-\Delta\right) d E^{ij} = 0\,.\ees
The second equation is simply the difference of
Eqs.~\eqref{eqEij}. This then permits to determine $d E^i$ and $d
E^{ij}$. Then, from the equality $d\Psi=d\Phi$ together
with~\eqref{dAdBz} we have
\bes{dC} && d F = d E' - \frac{1}{2}d B\,,\\ && d C = d F' +
2\mathcal{H} d F - \frac{1}{2}\mathcal{H}d B +
8\pi(\alpha+\varepsilon)\fond{\rho}\,a^2\,z\,,\ees
where $dF$ is a convenient intermediate notation. Thus, $d C$ is known
once $d F$ is known. Finally we combine the differences of~\eqref{eqX}
and~\eqref{Xdef} together with $d\Psi=d\Phi$ to obtain\,\footnote{The
  background equation $\mathcal{H}''-2\mathcal{H}\mathcal{H}' =
    - 4\pi (q+r)\,\mathcal{H}\fond{\rho}\,a^2$ is also used in this
  calculation. We recall from Eq.~\eqref{rhoM} that the matter density
  observed in cosmology is $\fond{\rho}_\text{M} =
    \frac{2\alpha\varepsilon}{\alpha+\varepsilon} \fond{\rho}$.}
\be{eqdPsi0} \Bigl(a\bigl(a\,d\Psi\bigr)'\Bigr)' = -4\pi
(q+r)\,a^4\fond{\rho}\,d\Psi\,,\ee
which can be transformed, \textit{via}
\be{eqdPsi} d\Psi = - \frac{1}{a}\bigl(a\,d F\bigr)' -
8\pi(\alpha+\varepsilon)\,a^2 \fond{\rho}\,z\,,\ee
into the following evolution equation which permits determining $d F$
and hence $d E$ and $d C$,
\be{eqdF} \Bigl(a\bigl(a\,d F\bigr)''\Bigr)' + 4\pi k\,(q+r)\bigl(a\,d
F\bigr)' = -8\pi k\,(\alpha+\varepsilon)\bigl[(a\,z')'+4\pi k (q+r)z
  \bigr]\,,\ee
where $k\equiv\,\fond{\rho}a^3$. The differences of gauge-invariant
velocity variables are also computed from
\bes{dVVi} d V' + \mathcal{H} d V + d \Phi &=& - 4\pi(1+\alpha)
\fond{\rho} a^2\,z\,,\\ d {V'}^i + \mathcal{H} d V^i &=& -
4\pi(1+\alpha) \fond{\rho} a^2\,z^i\,.\ees
Finally we conclude that all variables in our model can be fully and
consistently determined by solving linear evolution equations.

\bibliography{ListeRef.bib}

\end{document}